\documentclass{article}
\newcommand{\system}[1]{\textsc{#1}}

  \usepackage[preprint]{neurips_2026}

\usepackage[utf8]{inputenc} % allow utf-8 input
\usepackage[T1]{fontenc}    % use 8-bit T1 fonts
\usepackage{hyperref}       % hyperlinks
\usepackage{url}            % simple URL typesetting
\usepackage{booktabs}       % professional-quality tables
\usepackage{amsfonts}       % blackboard math symbols
\usepackage{nicefrac}       % compact symbols for 1/2, etc.
\usepackage{microtype}      % microtypography
\usepackage{xcolor}         % colors
\usepackage{comment}
\usepackage{graphicx}
\usepackage{graphicx}
\usepackage{subcaption}
\usepackage{float}
\usepackage{pifont}
\usepackage{tikz}
\usetikzlibrary{arrows.meta, positioning, shapes.geometric}
\usepackage{tikz}
\usepackage{xcolor}
\usetikzlibrary{arrows.meta, positioning, shapes.geometric, calc, fit, backgrounds}
\usepackage{longtable}
\definecolor{stageAudio}{HTML}{1F4E79}
\definecolor{stageAudioFill}{HTML}{DEEBF7}
\definecolor{stageTrans}{HTML}{8B3A00}
\definecolor{stageTransFill}{HTML}{FCE4D6}
\definecolor{stageQC}{HTML}{1F6B3A}
\definecolor{stageQCFill}{HTML}{E2F0D9}
\definecolor{endpointFill}{HTML}{FFF2CC}
\definecolor{endpointBorder}{HTML}{BF8F00}
\definecolor{vendorFill}{HTML}{F2F2F2}
\definecolor{vendorBorder}{HTML}{595959}
\definecolor{arrowGray}{HTML}{404040}
\definecolor{bandAudio}{HTML}{EAF2FB}
\definecolor{bandTrans}{HTML}{FBEEE4}
\definecolor{bandQC}{HTML}{ECF5E4}
\usepackage{xcolor}
\usetikzlibrary{arrows.meta, positioning, shapes.geometric, calc, fit, backgrounds}

\definecolor{onboardBorder}{HTML}{6B4E9B}
\definecolor{onboardFill}{HTML}{EDE5F7}
\definecolor{onboardBand}{HTML}{F5EFFC}
\definecolor{qcBorder}{HTML}{1F4E79}
\definecolor{qcFill}{HTML}{DEEBF7}
\definecolor{qcBand}{HTML}{EAF2FB}
\definecolor{decisionBorder}{HTML}{8B3A00}
\definecolor{decisionFill}{HTML}{FCE4D6}
\definecolor{acceptBorder}{HTML}{1F6B3A}
\definecolor{acceptFill}{HTML}{D6ECD9}
\definecolor{acceptBand}{HTML}{ECF5E4}
\definecolor{rejectBorder}{HTML}{A8322E}
\definecolor{rejectFill}{HTML}{F8D7D5}
\definecolor{rejectBand}{HTML}{FBE8E7}
\definecolor{arrowGray}{HTML}{404040}
\definecolor{expectedArrow}{HTML}{1F6B3A}
\definecolor{unexpectedArrow}{HTML}{A8322E}

\usepackage{tikz}
\usetikzlibrary{shapes.geometric, arrows.meta, positioning}
\usepackage{xcolor}

\definecolor{blueFill}{HTML}{E6F1FB}
\definecolor{blueStroke}{HTML}{185FA5}
\definecolor{blueText}{HTML}{0C447C}
\definecolor{tealFill}{HTML}{E1F5EE}
\definecolor{tealStroke}{HTML}{0F6E56}
\definecolor{tealText}{HTML}{085041}
\definecolor{purpleFill}{HTML}{EEEDFE}
\definecolor{purpleStroke}{HTML}{534AB7}
\definecolor{purpleText}{HTML}{3C3489}
\definecolor{amberFill}{HTML}{FAEEDA}
\definecolor{amberStroke}{HTML}{854F0B}
\definecolor{amberText}{HTML}{633806}
\definecolor{greenFill}{HTML}{EAF3DE}
\definecolor{greenStroke}{HTML}{3B6D11}
\definecolor{greenText}{HTML}{27500A}
\usepackage{xcolor}
\usepackage{longtable}
\usepackage{array}
\usepackage{booktabs}
\usepackage{colortbl}
\usepackage{tcolorbox}
\usepackage{hyperref}
\usepackage{enumitem}
\definecolor{primary}{RGB}{30, 90, 160}
\definecolor{tableHeader}{RGB}{30, 90, 160}
\definecolor{tableAlt}{RGB}{240, 245, 252}
\definecolor{accent}{RGB}{200, 70, 30}
\newtcolorbox{faqbox}[1]{
     colback=tableAlt,
     colframe=primary,
     fonttitle=\bfseries,
     title={#1},
     boxrule=0.5pt,
     arc=2pt,
     left=6pt, right=6pt, top=4pt, bottom=4pt,
     breakable
 }
\usepackage{xcolor}
\usepackage{longtable}
\usepackage{array}
\usepackage{booktabs}
\usepackage{colortbl}
\usepackage{tcolorbox}
\usepackage{hyperref}
\usepackage{enumitem}
\usepackage{seqsplit}

\definecolor{primary}{RGB}{30, 90, 160}
\definecolor{tableHeader}{RGB}{30, 90, 160}
\definecolor{tableAlt}{RGB}{240, 245, 252}
\definecolor{accent}{RGB}{200, 70, 30}
\definecolor{primary}{RGB}{30, 90, 160}
\definecolor{tableHeader}{RGB}{30, 90, 160}
\definecolor{tableAlt}{RGB}{240, 245, 252}
\definecolor{accent}{RGB}{200, 70, 30}

\providecommand{\fmt}[1]{\texttt{\small #1}}

\newcolumntype{N}{>{\raggedright\arraybackslash}p{2.5cm}}
\newcolumntype{C}{>{\centering\arraybackslash}p{0.7cm}}
\newcolumntype{D}{>{\raggedright\arraybackslash}p{1.4cm}}

\title{\system{VAANI}: Capturing the language landscape for an inclusive digital India}

\author{
\\
Sujith Pulikodan$^1$,
Abhayjeet Singh$^2$, Agneedh Basu$^1$, Nihar Desai$^1$, \\
Pavan Kumar J$^1$, Pranav D Bhat$^1$, Raghu Dharmaraju$^1$, Ritika Gupta$^2$, \\
Sathvik Udupa$^{2**}$, Saurabh Kumar$^2$,\\ Sumit Sharma$^2$, 
Visruth Sanka$^1$,\\ Dinesh Tewari$^3$, 
Harsh Dhand$^3$, Amrita Kamat$^3$, Sukhwinder Singh$^3$, \\
Shikhar Vashishth$^3$, Partha Talukdar$^3$, Raj Acharya$^{3*}$, \\
Prasanta Kumar Ghosh$^2$ \\
\\
$^1$AI \& Robotics Technology Park (ARTPARK), I-Hub @ IISc, Bangalore, India \\
$^2$Department of Electrical Engineering, Indian Institute of Science, Bangalore, India \\
$^3$Google DeepMind, Bangalore, India \\
\\
$^*$Currently at Quest Alliance \\
$^{**}$Work done while at Indian Institute of Science\\
spirelab.ee@iisc.ac.in, prasantg@iisc.ac.in
}
\begin{document}

\maketitle
\begin{abstract}
Voice-based technologies have the potential to bridge digital accessibility gaps; however, existing datasets fail to capture the linguistic and regional diversity of Indic languages. We present Project VAANI, a large-scale multimodal dataset designed to represent India’s linguistic landscape across 165 districts. Speech data is collected using image-based prompts to elicit spontaneous responses, while images are curated through a separate pipeline covering diverse themes across regions. The dataset undergoes a rigorous multi-stage quality control process, combining automated and manual evaluation to ensure high audio quality and transcription accuracy. We release approximately 289K images, 31,255 hours of speech, and 2,043 hours of transcribed audio spanning 105 languages from 28 states and 3 union territories. Many of these languages are represented at this scale for the first time, making VAANI a foundational resource for inclusive speech technology. The dataset enables the development of robust, multilingual, and multimodal models, and supports research in speech recognition, language understanding, and cross-modal learning for underrepresented languages.
\end{abstract}

%\section{Introduction}
%\label{sec:intro}

%\section{Introduction}
\section{Introduction}
\label{sec:intro}
India is one of the most linguistically diverse countries in the world. This linguistic diversity arises from a range of factors including historical and cultural influences, geographical dispersion, ethnic and social diversity, political-administrative needs and religious variations \cite{kidwai2019people}. Nearly 1.4 billion people in India communicate using a range of languages and dialects, often shaped by their regional and cultural context. According to the 2011 census \cite{census2011_languages}, after detailed linguistic scrutiny and rationalization, 1,369 mother tongues were identified and those spoken by over 10,000 individuals were grouped into 121 languages. These belong to five language families: Indo-European (23 languages, 78.05\% of the population), Dravidian (17 languages, 19.64\%), Austro-Asiatic (14 languages, 1.11\%), Tibeto-Burman (66 languages, 1.01\%), and Semito-Hamitic (1 language, <0.01\%). Among these, 22 languages are officially recognized under the Eighth Schedule of the Indian Constitution \cite{eighth_schedule}.

In multilingual contexts, voice-based technologies play a critical role in bridging accessibility gaps, particularly in domains such as education, governance, and health care \cite{kumar2012spoken}\cite{brewer2005case}. By reducing dependence on text-based literacy and supporting communication in native languages, speech interfaces promote greater inclusion in digital spaces \cite{wu2025speech}. Recent advances in AI, and particularly generative AI, have transformed how language is processed and generated~\cite{hagos2024recent,uddin2025critical}, with the field increasingly shifting toward multimodal models that jointly handle text, speech, and vision~\cite{caffagni2024revolution}. For these models to be inclusive, their training data must be representative of the linguistic and demographic diversity of the populations they aim to serve, capturing variations across languages, dialects, accents, and sociolinguistic groups that are systematically underrepresented in existing corpora ( Table \ref{tab:1}). Toward building such inclusive datasets, we open-source VAANI, a geo-centric, image-prompted multimodal corpus collected across 165 districts, spanning 105 languages, and comprising 31,255 hours of spontaneous speech, 2,043 hours of manual transcriptions, and 289,838 paired images from 158,441 speakers.

\begin{comment}
 Techniques like attention mechanisms\cite{vaswani2017attention} and diffusion models \cite{ho2020denoisingdiffusionprobabilisticmodels}, coupled with self-supervised pretraining followed by fine-tuning, have driven significant progress \cite{lee2023scalediversitycoefficientdata}. 
  However, these models rely heavily on high-quality and representative training data. However, for Indic languages, there remains a gap in representative datasets.
 
Language modeling efforts often treat “language” as a singular, monolithic label \cite{backus1999mixed}. In practice, spoken language varies significantly across regions, communities, education levels, and genders, even within the same official language \cite{shapiro2008language}\cite{sailaja2012indian}. Thus, datasets anchored only on language identifiers miss important regional and sociolinguistic variations. To develop inclusive and robust models, we need data that reflects both language-anchored and region-anchored variations in speech patterns.
\end{comment}

%Speech data collection methods are typically classified as:

    %Read speech, where speakers read predefined scripts; it is easier to organize and %annotate but lacks natural variability.

    %Spontaneous speech, where speakers respond freely to stimuli, capturing more %realistic and diverse speech but requiring more effort to annotate.

\section{Related Work}
\label{sec:related}

Primary data collection for speech technology typically follows two main strategies: \textit{read speech}, where speakers recite predetermined text, and \textit{spontaneous speech}, where speakers produce unscripted utterances in response to stimuli. Read speech simplifies organization and annotation since the content is controlled and consistent, making it well-suited for initial dataset creation where linguistic clarity is key. Spontaneous speech, by contrast, captures more authentic language use and richer linguistic diversity, but is more resource-intensive to transcribe and annotate, since it must reflect real-world communication patterns rather than scripted content.
\begin{table*}[t]
  \centering
  \caption{Comparison of Indic speech datasets. (Mixed refers to corpora combining read and spontaneous speech. NA indicates data not reported by the original authors.)}
  \begin{tabular}{@{}p{5.0cm}p{2.0cm}p{2.0cm}p{1.0cm}p{1.8cm}@{}}
  \toprule
  \textbf{Dataset} & \textbf{Collection Method} & \textbf{Duration (Hours)} & \textbf{\# Languages} & \textbf{\# Speakers} \\
\midrule
IndicVoice\cite{C4}        & Mixed & 23.7K & 22 &51K  \\
RESPIN\cite{kumarrespin}           & Read  & 10,416  &9  &18K+ \\
SYSPIN\cite{abhayjeet2025syspin}           &Read    &900      &9   &18  \\
Shrutilipi \cite{bhogale2023effectiveness}       & Read  & 6457  & 12 &NA  \\
Kathbath \cite{javed2023indicsuperb}         & Read  & 1684  & 12 &1217  \\
Spring-inx\cite{gangwar2023spring}        & Mixed & 2005  & 10 &7609  \\
FLEURS \cite{conneau2023fleurs}           & Read & 163   & 13 &39  \\
CommonVoice \cite{ardila2019common}       & Read  & 373   & 10 &NA  \\
IndicTTS\cite{baby2016resources}          & Read  & 225   & 13 &26  \\
NPTEL\cite{bhogale2023vistaar}             & Read  & 857   & 8  &NA  \\
Gram Vaani\cite{bhanushali22_interspeech}        & Spontaneous & 1108 & 1 &NA  \\
IISc-MILE\cite{pilar2022subword}         & Read  & 497   & 2  &1446  \\
MUCS \cite{Diwan_2021}              & Mixed & 600    & 7  &310  \\
E\&NE languages\cite{basu2021multilingual}   & Mixed & 19.75 & 4  &NA  \\
NISP \cite{kalluri2021nisp}             & Mixed & 56.86 & 6  &345  \\
CMS\cite{he2020open}               & Read  & 35    & 6  &243  \\
IndicSpeech \cite{srivastava2020indicspeech}      & Read  & 24    & 3  &3  \\
MSR Challenge \cite{srivastava2018interspeech}     & Mixed & 150   & 3  &1286  \\
IIITH-ILSC\cite{vuddagiri2018iiith}        & Mixed & 103.5 & 23 &1150  \\
IIITH-ISD\cite{prahallad2012iiit}         & Read  & 11    & 7  &35  \\
\midrule
\textbf{VAANI}            & Spontaneous & \textbf{31255}   & \textbf{105} &\textbf{158K}  \\
\bottomrule
\end{tabular}

\label{tab:1}
\end{table*}

Efforts to collect speech data for Indic languages (Table~\ref{tab:1}) have predominantly focused on the 22 languages officially recognized under the Eighth Schedule of the Indian Constitution. As Table~\ref{tab:1} shows, no existing open Indic corpus covers more than 23 languages, leaving the majority of India's 100+ mother tongues~\cite{census2011_languages} without representation in publicly available speech resources. Beyond language count, language modeling efforts often treat "language" as a singular, monolithic label~\cite{backus1999mixed}. In practice, spoken language varies significantly across regions, communities, education levels, and genders, even within the same official language~\cite{shapiro2008language,sailaja2012indian}. Datasets anchored only on language identifiers therefore miss important regional and sociolinguistic variation, particularly in a country as linguistically and geographically diverse as India. Building inclusive and robust models requires data that reflects both language-anchored and region-anchored variation in speech patterns.

Finally, AI systems are increasingly multimodal, jointly modeling speech, vision, and text. Yet most publicly available Indic speech corpora pair audio only with text transcriptions, lacking the visual grounding needed to train and evaluate multimodal models tailored to Indic contexts.
In summary, existing Indic speech corpora are limited along three dimensions: narrow language coverage, language-anchored rather than region-anchored sampling, and a speech–text-only modality structure. VAANI addresses all three simultaneously by adopting a geo-centric data collection strategy across 165 districts, capturing 105 languages, and pairing every audio recording with the visual prompt that elicited it, yielding aligned image–speech–text triplets that support multimodal model development for Indic language.
\section{The Dataset}
\label{sec:dataset}

%\todo{PT: we can also make this the third section -- introduce the dataset first and %then talk about how we built it?}
VAANI is an India-representative multimodal, multilingual dataset. It contains audio recordings collected along with the images shown to the speakers, and where available, the corresponding transcriptions. Each speaker was presented with an image and asked to state what comes to their mind in their own words, allowing for the creation of aligned image–speech–text triplets. The dataset includes data collected from 165 districts across the country, covering 105 languages. It comprises  24,009,427 audio segments collected from 158,441 speakers recorded in response to 289,838 images, with a total audio duration of approximately 31,255 hours. From this collection, 2,043 hours of audio have been manually transcribed, with transcription data distributed nearly evenly across the 165 districts, ensuring balanced geographic representation. The data has been collected from 165 districts across 28 states and 3 union territories in India.

\begin{figure}[h]
\centering
\begin{minipage}[c]{0.55\textwidth}
    \centering
    \includegraphics[width=\linewidth]{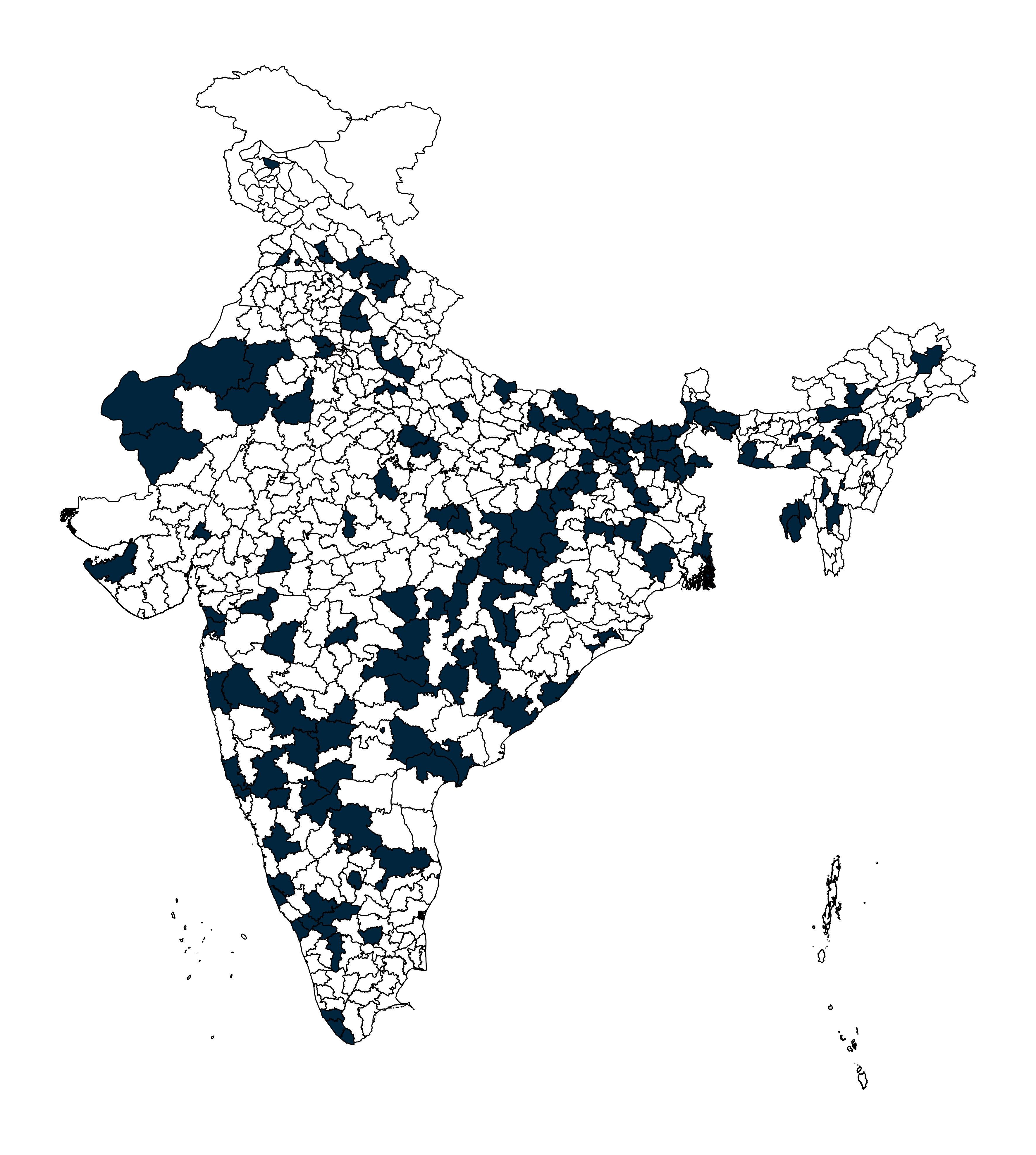}
    \subcaption{Geographic coverage across Indian districts.}
    \label{fig:coverage_map}
\end{minipage}\hfill
\begin{minipage}[c]{0.42\textwidth}
    \centering
    \begin{tikzpicture}[
        node distance=0.35cm,
        every node/.style={font=\sffamily\scriptsize},
        block/.style={
            rectangle, rounded corners=3pt, draw, line width=0.4pt,
            text width=3.4cm, align=center, inner sep=4pt, minimum height=0.7cm
        },
        arrow/.style={-{Stealth[length=2mm]}, line width=0.5pt}
    ]
    \node[block, fill=blueFill, draw=blueStroke] (img)
        {{\color{blueText}\textbf{Image data collection}}\\{\color{blueText}\tiny Capture raw images}};
    \node[block, fill=blueFill, draw=blueStroke, below=of img] (imgqc)
        {{\color{blueText}\textbf{Quality check}}\\{\color{blueText}\tiny Validate image quality}};
    \node[block, fill=tealFill, draw=tealStroke, below=of imgqc] (speech)
        {{\color{tealText}\textbf{Speech data collection}}\\{\color{tealText}\tiny Record audio samples}};
    \node[block, fill=tealFill, draw=tealStroke, below=of speech] (speechqc)
        {{\color{tealText}\textbf{Quality check}}\\{\color{tealText}\tiny Validate audio quality}};
    \node[block, fill=purpleFill, draw=purpleStroke, below=of speechqc] (tsel)
        {{\color{purpleText}\textbf{Transcription selection}}\\{\color{purpleText}\tiny Choose samples to transcribe}};
    \node[block, fill=purpleFill, draw=purpleStroke, below=of tsel] (tproc)
        {{\color{purpleText}\textbf{Transcription process}}\\{\color{purpleText}\tiny  Transcribe the data}};
    \node[block, fill=amberFill, draw=amberStroke, below=of tproc] (finalqc)
        {{\color{amberText}\textbf{Quality check}}\\{\color{amberText}\tiny Automated and Manual check}};
    \node[block, fill=greenFill, draw=greenStroke, below=of finalqc] (release)
        {{\color{greenText}\textbf{Open source release}}\\{\color{greenText}\tiny Publish dataset}};
    \draw[arrow] (img) -- (imgqc);
    \draw[arrow] (imgqc) -- (speech);
    \draw[arrow] (speech) -- (speechqc);
    \draw[arrow] (speechqc) -- (tsel);
    \draw[arrow] (tsel) -- (tproc);
    \draw[arrow] (tproc) -- (finalqc);
    \draw[arrow] (finalqc) -- (release);
    \end{tikzpicture}
    \subcaption{Data collection and processing pipeline.}
    \label{fig:data-pipeline}
\end{minipage}
\caption{Geographic coverage of VAANI across Indian districts (left), and the data collection pipeline (right)}
\label{fig:coverage_overview}
\end{figure}

Districts were selected to maximize coverage of the languages reported in the 2011 Census of India \cite{census2011_languages} along with broad geographic representation. For the majority of districts, approximately 200 hours of audio were collected, of which 5–10\% was manually transcribed. To capture intra-district linguistic variation, recordings were gathered from multiple locations within each district. The 105 languages (Table \ref{tab:languages}) span four language families: Indo-Aryan (e.g., Hindi, Bengali, Marathi, Maithili, Bhojpuri, Chakma), Dravidian (Tamil, Telugu, Kannada, Malayalam, Tulu), Tibeto-Burman (Mizo, Garo, Wancho, Idu Mishmi, Lepcha), and Austroasiatic (Santali). Many of these — particularly Tibeto-Burman and Austroasiatic varieties — appear at this scale in an open speech dataset for the first time.
The language-wise distribution skews toward a few mainstream languages, but this should be read alongside the geographic spread: recordings come from a wide range of districts, each contributing regional dialectal and cultural variation that monolithic "language" labels would otherwise obscure. Capping transcription at roughly 5–10\% of the audio is a deliberate design choice that balances annotation cost against collection scale; the untranscribed bulk supports unsupervised and image-aligned pretraining, while the transcribed subset enables supervised fine-tuning, mirroring the two-stage paradigm of modern speech foundation models.

\begin{table}[h]
\centering
\caption{Per-segment metadata fields released with VAANI.}
\small
\begin{tabular}{@{}lp{8cm}@{}}
\toprule
\textbf{Field} & \textbf{Description} \\
\midrule
State, District, Pincode & Geographic localization at multiple granularities \\
Language & Speaker's language for the recording \\
Languages Known & Speaker's multilingual repertoire \\
Speaker ID & Anonymous unique speaker identifier \\
Speaker Image Hash & Unique speaker--image pairing identifier \\
Reference Image Name & The visual prompt shown to the speaker \\
Gender & Demographic attributes \\
Stay (years) & Residential history (location, duration) \\
\bottomrule
\end{tabular}
\label{tab:metadata}
\end{table}

During data collection, additional metadata fields were captured to support quality control, speaker validation, and inclusive representation across demographic groups. These include age, education level, self-reported socio-economic status, and recording device information (e.g., phone brand and model). Collectively, these metadata attributes enable robust categorization and significantly enhance the dataset’s utility for developing regionally and linguistically inclusive speech technologies.

Following a rigorous curation process, the dataset has been open-sourced via Hugging Face under CC-BY-4.0 , with the option to download data at the district level. We also facilitate the downloading of transcribed data at the language level.

\section{Data Collection}
\label{sec:data_collection}

We aimed to build a dataset that prioritizes inclusivity of people rather than focusing solely on languages. Our data collection process was designed to encourage natural and unbiased expression, ensuring a more authentic representation of how people communicate. Instead of adopting a language-centric approach, we followed a district-centric strategy to capture speech data from individuals, representing a wide range of demographics. This included considerations such as gender, socioeconomic background, education, and age. Organizing data collection at the district level enabled an efficient and scalable process. The project was executed in two phases. In Phase~1, we successfully covered 80 districts across 12 states. In Phase~2, we covered an additional 85 districts across 23 states and 3 Union Territories, ensuring a broad representation of the country's linguistic and social diversity (Figure~\ref{fig:coverage_map}). This approach enables a more comprehensive understanding of speech patterns and language use across diverse communities. 

For data collection, we used carefully selected images as visual stimuli. When presented with an image, the speaker was encouraged to say whatever came to mind, fostering natural and varied responses. Image prompts offer three advantages over alternatives: fewer structural constraints than text, accessibility for non-literate speakers and unscripted languages (e.g., Paniya, Chakma, Wancho), and reduced social-desirability bias compared to conversational elicitation.

%We collected  approximately 1700 to 2000 images from each district, apart from 1500 %images which are not tagged to any district (generic images). This makes for a rich %image dataset, which later acts as a prompt for spontaneous speech. 

%\subsection{Collection Strategy}

%\begin{figure*}[t]
 % \includegraphics[width=\textwidth]{figs/VAANI.png}
 % \caption{Automated QC-Process flow}
%\end{figure*}
\begin{comment}
\subsection{Collection goals}
The districts were selected based on linguistic diversity, with the goal of covering 80 districts in Phase 1 and  85 districts in Phase 2, across 25 states in India. From each district, our goal was to open source around 200 hours of audio data, with 10\% of it manually transcribed. Comprehensive metadata was collected for each speaker, including age, gender, languages spoken, education, socio-economic background, location, and duration of stay in that location. To ensure ethical data use, each participant provided a signed consent form, transferring ownership of the data and enabling its open-source distribution without restrictions.
.
\end{comment}

%\subsection{Collection Process }
The collection process (Figure \ref{fig:data-pipeline}) is operationally challenging, resource intensive, and time-consuming. To address this, we enlisted the support of vendors to manage the process effectively. The vendors established an extensive network of coordinators in each district to facilitate data collection. For audio collection and transcription, we collaborated with three vendors, Shaip\cite{shaip2025}, Megdap\cite{megdap2025}, and Karya\cite{karya2025} each tasked with providing high-quality audio data, and transcriptions of 10\% of those audios, that pass the quality evaluations. We also engaged GTS\cite{gtsai2025} as the vendor to collect images from each district. 
%\subsubsection{}
\subsection{Image Acquisition}

To facilitate the collection process, we gathered 1,700 to 2,000 images per district, consisting of both district-specific and general images. These images were newly captured for this project with assistance from external vendors. The topics for each district were identified based on local context and shared with the vendors. They were instructed to deliver images in .jpg format, with a resolution of 640x400 pixels and a file size of under 500 KB.

The image collection process adhered to a set of strict specifications to ensure quality, relevance, and ethical standards. All images were required to be physically captured and not sourced from online or third-party media. Each image needed to be clear, well-focused, and free from any blurring or visual distortions. The object relevant to each category had to be prominently visible without any obstructions. Furthermore, images were required to be unique for every district and category, with no duplication permitted across different regions or categories. To maintain privacy and compliance, images were not allowed to include personally identifiable information such as people, recognizable logos, or privately owned objects. Importantly, the images needed to depict content that could be meaningfully described, as they were intended to prompt spoken responses from participants. Finally, the date of image capture was mandated to be no earlier than July 1, 2023, ensuring the recency and relevance of the content.

To ensure quality, we developed a quality check workflow to evaluate the images delivered by vendors. This process included both automated validation and a human-in-the-loop review. Each image underwent a series of automated and manual checks to verify quality, uniqueness, and compliance with provided guidelines. Only those images that satisfied all the defined criteria were accepted and subsequently used for data collection.

\subsection{Speech Acquisition}
To facilitate high-quality data collection, we established comprehensive audio and transcription specifications that were shared with the vendors involved in the project. These specifications defined both the technical format of the recordings and the style of data collection to ensure consistency, clarity, and linguistic diversity. Vendors were also given clear instructions regarding the style of data collection. The speech had to be spontaneous, with utterances spoken in response to visual prompts. Participants were shown images selected randomly from a curated pool, and for each image, they were expected to provide 10–20 seconds of effective speech. Instructions emphasized the importance of variety in the spoken responses to ensure linguistic richness and diversity. Speaker selection followed  criteria to ensure nativity, demographic diversity, and balanced representation. Participants were required to be between the ages of 20 and 70, with uniform distribution across this age range. Each district had to include at least 800 speakers, with gender balance maintained. No individual speaker was permitted to contribute more than 15 minutes of effective speech. All speakers were required to be native residents of the pincode recorded in the metadata, verified using identity documents. Participants were encouraged to speak in the language or dialect they typically use at home with family members. 

To collect data from speakers, the vendors utilized  mobile application along with a web application for quality control. Coordinators, selected and onboarded from each district, traveled to assigned locations within their districts. They onboarded speakers while ensuring diversity across gender, age groups, education levels, and socio-economic backgrounds. Speakers were registered on the platform with the assistance of coordinators. Initially, they were shown sample images and asked to speak about them. The sample data was then validated by the vendor's quality control teams. If the data met quality standards, the speaker was officially onboarded to the platform. After onboarding, speakers were presented with a set of images, including district-specific and generic ones, and were asked to record audio. 
Once the recordings were completed, the application provided an option to submit the data, which was then processed further. The data collected from users undergoes an initial quality check by the vendors to ensure its integrity. This includes verifying the audio quality, as well as the completeness and accuracy of the metadata. To ensure geographical coverage, pincode information is captured during user onboarding, and the GPS location of the mobile device is used for validation. Once the data is collected and validated by the vendors, it is submitted for further processing.
\subsection{Transcription}
The pipeline is designed to streamline the selection of audio segments for transcription at the district level, focusing on metadata utilization, speaker diversity, and balanced language representation. By incorporating statistical analysis, random sampling, and transcript-based strategies, it ensures fair and comprehensive audio file selection. Segment selection follows systematic approaches that prioritize linguistic diversity and equitable speaker representation while avoiding over-representation of dominant demographics. Transcript-based selection and threshold checks further enhance balance across multiple languages and districts. The iterative process dynamically adjusts to achieve target duration thresholds without compromising diversity. District-level analysis identifies data coverage gaps through calculated selection rates, ensuring uniform data collection and avoiding the neglect of underrepresented regions. Metadata management is a key aspect, with selected files recorded alongside attributes such as filename, district, language, duration, and selection criteria. Segments for transcription are selected from the audio data  based on this criteria and sent back to the vendors. To maintain linguistic and contextual accuracy, specifications require transcribers to be from the same district where the audio data was collected. Vendors recruited and trained transcribers, post which they transcribe the selected audio segments on the basis of guidelines and specifications. Once transcription is completed and undergoes quality control, the transcribed data is submitted  for further processing.

\section{Quality Control}
The dataset underwent a comprehensive quality assurance process to ensure its reliability and accuracy across all components, including audio recordings, metadata, transcriptions, and data formatting. The quality evaluation pipeline was meticulously designed to identify and resolve any issues that could compromise the dataset's integrity. The process incorporated both automated and manual checks. Automated checks evaluated the quality of audio, metadata, and transcriptions. Manual checks were conducted by trained evaluators from the corresponding districts to address finer nuances that automated tools don't account for. Local evaluators played a crucial role in maintaining contextual and linguistic integrity. The vendors submitted the data in organized batches through repositories hosted on Google Cloud Platform (GCP). Each batch underwent a final review to confirm compliance with quality standards before being integrated into the larger dataset for further analysis and processing.

\begin{figure}[htbp]
    \centering
    \resizebox{\textwidth}{!}{%
\begin{tikzpicture}[
    font=\sffamily\bfseries,
    >={Stealth[length=2.6mm, width=2.4mm]},
    node distance=6mm and 7mm,
    stepAudio/.style={rectangle, rounded corners=3pt, draw=stageAudio, line width=1.0pt,
        fill=stageAudioFill, minimum width=2.6cm, minimum height=1.05cm, align=center,
        font=\normalsize\bfseries, text=stageAudio!30!black},
    stepTrans/.style={rectangle, rounded corners=3pt, draw=stageTrans, line width=1.0pt,
        fill=stageTransFill, minimum width=2.9cm, minimum height=1.05cm, align=center,
        font=\normalsize\bfseries, text=stageTrans!40!black},
    stepQC/.style={rectangle, rounded corners=3pt, draw=stageQC, line width=1.0pt,
        fill=stageQCFill, minimum width=2.6cm, minimum height=1.05cm, align=center,
        font=\normalsize\bfseries, text=stageQC!30!black},
    endpoint/.style={rectangle, rounded corners=3pt, draw=endpointBorder, line width=1.1pt,
        fill=endpointFill, minimum width=2.6cm, minimum height=1.05cm, align=center,
        font=\normalsize\bfseries, text=endpointBorder!30!black},
    vendor/.style={rectangle, rounded corners=4pt, draw=vendorBorder, line width=1.0pt,
        fill=vendorFill, minimum width=2.0cm, minimum height=1.05cm, align=center,
        font=\normalsize\bfseries, text=black},
    format/.style={rectangle, rounded corners=3pt, draw=black!70, line width=1.0pt,
        fill=black!10, minimum width=3.3cm, minimum height=1.05cm, align=center,
        font=\normalsize\bfseries, text=black},
    flow/.style={-{Stealth[length=2.6mm, width=2.4mm]}, draw=arrowGray, line width=1.0pt},
    flowThick/.style={-{Stealth[length=2.8mm, width=2.6mm]}, draw=arrowGray, line width=1.2pt},
    vendorLink/.style={dotted, line width=1.0pt, draw=black!60},
    edgelabel/.style={font=\small\bfseries, text=black, inner sep=2pt,
        fill=white, fill opacity=0.95, text opacity=1},
    stagetag/.style={font=\normalsize\bfseries, inner sep=3pt, rounded corners=2pt}
]

% ROW 1 - Audio Collection
\node[vendor] (v1) at (0,0) {Vendors};
\node[stepAudio, right=of v1] (a1) {Level 1\\Check};
\node[stepAudio, right=of a1] (a2) {Level 2\\Check};
\node[stepAudio, right=of a2] (a3) {Level 3\\Check};
\node[stepAudio, right=of a3] (a4) {Manual\\QC};
\node[endpoint, right=of a4]  (a5) {Accepted\\Data};
\draw[flow] (v1)--(a1); \draw[flow] (a1)--(a2); \draw[flow] (a2)--(a3);
\draw[flow] (a3)--(a4); \draw[flow] (a4)--(a5);

% ROW 2 - Transcription Selection + Data Formatting
\node[stepTrans, below=14mm of a3] (t1) {Transcription\\Segment Selection};
\node[vendor, left=40mm of t1] (v2) {Vendors};
% Place fmt to the right of a5, at row 2's vertical level
\coordinate (fmtAnchor) at ([xshift=14mm]a5.east);
\node[format] (fmt) at (fmtAnchor |- t1) {Data Formatting\\for Open-Sourcing};

\draw[flow] (a5.south) |- (t1.east);
\draw[flow] (t1.west) -- node[edgelabel, above, align=center]{Segments for\\transcription}(v2.east);

% ROW 3 - Transcription QC, aligned with v1 of row 1
\node[vendor] (v3) at ($(v1)+(0,-5.5)$) {Vendors};
\node[stepQC, right=18mm of v3] (q1) {Level 1\\Check};
\node[stepQC, right=of q1] (q2) {Level 2\\Check};
\node[stepQC, right=of q2] (q3) {QC Files\\Selection};
\node[stepQC, right=of q3] (q4) {Manual\\QC};
\node[endpoint] (q5) at (fmt |- v3) {Accepted\\Data};
\draw[flow] (v3)-- node[edgelabel, above, align=center]{Transcribed\\audio}(q1);
\draw[flow] (q1)--(q2); \draw[flow] (q2)--(q3);
\draw[flow] (q3)--(q4); \draw[flow] (q4)--(q5);

% CONVERGENCE into fmt
%   - a5 exits east, travels right, drops down, enters fmt at TOP CENTER
%   - q5 (now horizontally aligned with fmt) goes straight up into BOTTOM CENTER
\draw[flow] (a5.east) -| (fmt.north);
\draw[flow] (q5.north) -- (fmt.south);

% DOTTED CONNECTIONS BETWEEN VENDORS (no arrows)
% Routes through a shared vertical rail on the left side of v1/v3
\coordinate (vendorRail) at ([xshift=-6mm]v1.west);
\draw[vendorLink] (v1.west) -| (vendorRail) |- (v2.west);
\draw[vendorLink] (v2.west) -| (vendorRail) |- (v3.west);

% BANDS
\begin{scope}[on background layer]
    \node[fill=bandAudio, rounded corners=6pt, fit=(v1)(a5), inner xsep=4mm, inner ysep=3mm] (bandA) {};
    \node[fill=bandTrans, rounded corners=6pt, fit=(v2)(t1), inner xsep=4mm, inner ysep=3mm] (bandT) {};
    \node[fill=bandQC, rounded corners=6pt, fit=(v3)(q5), inner xsep=4mm, inner ysep=3mm] (bandQ) {};
\end{scope}

% STAGE TAGS
\node[stagetag, fill=stageAudio, text=white, anchor=south west]
    at ([xshift=4mm, yshift=0.5mm]bandA.north west){STAGE 1 \textbf{\textbar}\ Audio Collection};
\node[stagetag, fill=stageTrans, text=white, anchor=south west]
    at ([xshift=4mm, yshift=0.5mm]bandT.north west){STAGE 2 \textbf{\textbar}\ Transcription Selection};
\node[stagetag, fill=stageQC, text=white, anchor=south west]
    at ([xshift=4mm, yshift=0.5mm]bandQ.north west){STAGE 3 \textbf{\textbar}\ Transcription QC};

\end{tikzpicture}
    }
    \caption{VAANI data collection and quality-control pipeline. Dotted lines indicate the same vendor pool operating across stages.}
    \label{fig:pipeline}
\end{figure}

\subsection{Automated Audio Quality Control}

In this process, the quality of audio, file format, audio specifications, and metadata is evaluated at multiple levels depending on the complexity of the workflow. This multi-stage validation approach enables timely feedback to vendors, allowing them to take corrective actions and resubmit data as needed.

In \textbf{Level 1}, the workflow performs a comprehensive quality assessment to ensure accuracy and consistency across metadata and audio files. It begins with structural validation, where metadata files are checked against a strict format. File naming and path validation is also enforced to ensure
consistency with predefined standards. Data consistency checks are carried out by cross-referencing metadata with actual data entries, identifying missing or duplicate speaker and utterance IDs, and ensuring that references to associated files are accurate. The workflow also verifies the presence of mandatory consent forms and supporting documents, flagging any missing items as critical issues. Geographical and contextual validation is performed by cross-verifying metadata fields such as district and state, ensuring alignment with associated audio and image files.

In \textbf{Level 2}, the workflow conducts a deeper validation of both audio data and metadata to ensure dataset integrity. Audio files are first evaluated for basic technical properties, including a 16 kHz sample rate, mono channel configuration, and 16-bit precision. File integrity checks identify duplicate filenames across current and previous batches and ensure adherence to naming conventions. Segment-level validation examines issues such as overlapping durations, negative durations, excessively short segments (less than 0.5 seconds), and speakers exceeding the maximum allowed duration of 15 minutes. Silence analysis using Voice Activity Detectors (VAD) is performed to assess clarity, flagging segments with excessive silence at the beginning or end (greater than 0.3 seconds) or within segments (greater than 1 second). Metadata validation in this stage includes demographic checks to ensure logical and geographical consistency. For example, pincodes must correspond to the specified state and the reported duration of residence cannot exceed the age of the speaker. Additionally, the language of the recording must match the languages reported by the speaker. Consistency across batches is also examined to detect conflicting demographic details for the same speaker. Speaker validation identifies incomplete or inconsistent records, flagging speakers with fewer than 10 valid audio files or conflicting metadata for exclusion. Comprehensive cross-validation ensures proper alignment between metadata and audio, highlighting mismatches in filenames, segment durations, or demographic information. These checks collectively maintain high standards of quality and consistency, enabling reliable downstream processing and analysis.

In \textbf{Level 3}, the quality of individual audio segments is evaluated using the Signal-to-Noise Ratio (SNR). This metric measures the strength of the desired audio signal relative to background noise and serves as an indicator of recording clarity. For each segment, the computed SNR is compared against a predefined threshold to ensure acceptable quality. Segments that do not meet this threshold are flagged for manual review, ensuring that only high-quality recordings are retained for further use.

\begin{comment}
\subsubsection{Sample Selection for Audio Manual QC process}

The files flagged during the Level 3 check are selected for manual review. Additionally, we ensure that at least one segment from each speaker is included. These segments are randomly sampled so that 10\% of the data that passed both the Level 1 and Level 2 checks also undergoes manual quality control.
\end{comment}
\subsection{Automated Transcription Quality Control}

The automated transcription check consists of two levels of check, and in  \textbf{Level 1}, the transcription data undergo rigorous quality checks to ensure accuracy and consistency. Each segment and its associated metadata are scrutinized for potential errors. The checks validate that segment names and transcriber IDs are correctly assigned, and the segment duration aligns with the number of words, flagging segments where the word count falls below a defined threshold. Transcriptions are evaluated for the presence of invalid elements such as numbers, incomplete sentences, or missing punctuation like open or unclosed brackets. Additionally, checks ensure that non-native characters are not used, and any missing or mismatched segments between the transcription files and the original audio data are identified. The transcription metadata undergoes a parallel validation process.  Invalid symbols in language-related fields and discrepancies in transcriber information, such as mismatched or missing demographic details, are also highlighted. Finally, checks confirm the proper formatting of metadata files, ensuring no blank lines, incorrect colon usage, or repeated entries within a batch. 

In \textbf{ Level 2}, the transcription data undergoes advanced quality checks to ensure linguistic and structural consistency, leveraging both statistical and logical evaluations. Each segment is analyzed for adherence to defined thresholds, such as Language Model log-likelihood, where transcriptions must exhibit linguistic plausibility within acceptable bounds. The Word Error Rate (WER) is also assessed, with segments exceeding predefined thresholds flagged for review.  Language consistency is scrutinized by cross-checking the transcription language with the expected language for the segment. A mismatch triggers an error for correction. The geographic relevance of the transcription is also validated, comparing pincode information against the expected region. A significant difference between the pincode in the transcription metadata and the designated area triggers a warning. These checks collectively ensure that the transcriptions are linguistically accurate, structurally coherent, and contextually relevant, contributing to a high-quality data set for downstream applications.
\begin{figure}[htbp]
\centering
\resizebox{\textwidth}{!}{%
\begin{tikzpicture}[
    font=\sffamily\bfseries,
    >={Stealth[length=2.6mm, width=2.4mm]},
    node distance=8mm and 9mm,
    onboardStep/.style={ellipse, draw=onboardBorder, line width=1.0pt,
        fill=onboardFill, minimum width=4.0cm, minimum height=1.6cm,
        align=center, font=\normalsize\bfseries, text=onboardBorder!40!black},
    qcStep/.style={rectangle, rounded corners=4pt, draw=qcBorder, line width=1.0pt,
        fill=qcFill, minimum width=2.7cm, minimum height=1.15cm,
        align=center, font=\normalsize\bfseries, text=qcBorder!30!black},
    qcStepWide/.style={rectangle, rounded corners=4pt, draw=qcBorder, line width=1.0pt,
        fill=qcFill, minimum width=3.2cm, minimum height=1.15cm,
        align=center, font=\normalsize\bfseries, text=qcBorder!30!black},
    decision/.style={rectangle, rounded corners=4pt, draw=decisionBorder, line width=1.2pt,
        fill=decisionFill, minimum width=3.2cm, minimum height=1.4cm,
        align=center, font=\normalsize\bfseries, text=decisionBorder!40!black},
    expectedBranch/.style={rectangle, rounded corners=4pt, draw=acceptBorder, line width=0.9pt,
        fill=acceptFill, minimum width=2.5cm, minimum height=1.1cm,
        align=center, font=\normalsize\bfseries, text=acceptBorder!40!black},
    unexpectedBranch/.style={rectangle, rounded corners=4pt, draw=rejectBorder, line width=0.9pt,
        fill=rejectFill, minimum width=2.5cm, minimum height=1.1cm,
        align=center, font=\normalsize\bfseries, text=rejectBorder!40!black},
    acceptNode/.style={rectangle, rounded corners=4pt, draw=acceptBorder, line width=1.2pt,
        fill=acceptFill, minimum width=2.3cm, minimum height=1.15cm,
        align=center, font=\normalsize\bfseries, text=acceptBorder!30!black},
    rejectNode/.style={rectangle, rounded corners=4pt, draw=rejectBorder, line width=1.2pt,
        fill=rejectFill, minimum width=2.3cm, minimum height=1.15cm,
        align=center, font=\normalsize\bfseries, text=rejectBorder!30!black},
    outcomeNode/.style={rectangle, rounded corners=4pt, draw=black!75, line width=1.2pt,
        fill=black!88, text=white, minimum width=2.5cm, minimum height=1.15cm,
        align=center, font=\normalsize\bfseries},
    flow/.style={-{Stealth[length=2.6mm, width=2.4mm]}, draw=arrowGray, line width=1.0pt},
    flowAccept/.style={-{Stealth[length=2.8mm, width=2.6mm]}, draw=expectedArrow, line width=1.1pt},
    flowReject/.style={-{Stealth[length=2.8mm, width=2.6mm]}, draw=unexpectedArrow, line width=1.1pt},
    edgelabel/.style={font=\small\bfseries, text=black, inner sep=2pt,
        fill=white, fill opacity=0.95, text opacity=1},
    stagetag/.style={font=\normalsize\bfseries, inner sep=3.5pt, rounded corners=2pt}
]

% Phase 1 - Onboarding
\node[onboardStep] (o1) {Send Welcome Kit\\with Legal Document};
\node[onboardStep, below=of o1] (o2) {Acceptance of T\&C\\and SoW};
\node[onboardStep, below=of o2] (o3) {Train Freelancers};
\draw[flow] (o1) -- (o2);
\draw[flow] (o2) -- (o3);

% Phase 2 - QC Workflow
\node[qcStepWide, right=22mm of o3] (assign_qc) {Assign Single-Audio QC\\(generic \& district checks)};
\node[qcStep, above=20mm of assign_qc] (assign_file) {Assign 1 File\\to 1 Freelancer};
\draw[flow] (o3) -- (assign_qc);
\draw[flow] (assign_qc) -- (assign_file);

\node[expectedBranch, above right=6mm and 14mm of assign_file] (expected) {Expected\\Response};
\node[qcStep, right=12mm of expected] (sample) {Review 10\%\\(sample basis)};
\node[unexpectedBranch, below right=6mm and 14mm of assign_file] (unexpected) {Unexpected\\Response};
\node[qcStep, right=12mm of unexpected] (allfiles) {Review\\All Files};
\node[decision, right=14mm of assign_file, xshift=46mm] (decision) {Follow\\Decision-Making};

\draw[flowAccept] (assign_file.east) to[out=30, in=180] (expected.west);
\draw[flowAccept] (expected) -- (sample);
\draw[flowAccept] (sample.east) to[out=0, in=150] (decision.north west);
\draw[flowReject] (assign_file.east) to[out=-30, in=180] (unexpected.west);
\draw[flowReject] (unexpected) -- (allfiles);
\draw[flowReject] (allfiles.east) to[out=0, in=-150] (decision.south west);

% Outcomes
\node[acceptNode, above right=4mm and 16mm of decision] (accepted) {Accepted};
\node[outcomeNode, right=of accepted] (speechAcc) {Speech Data\\Accepted};
\node[rejectNode, below right=4mm and 16mm of decision] (rejected) {Rejected};
\node[outcomeNode, right=of rejected] (speechRej) {Speech Data\\Rejected};

\draw[flowAccept] (decision.north east) to[out=30, in=180] (accepted.west);
\draw[flowAccept] (accepted) -- (speechAcc);
\draw[flowReject] (decision.south east) to[out=-30, in=180] (rejected.west);
\draw[flowReject] (rejected) -- (speechRej);

% Background bands
\begin{scope}[on background layer]
    \node[fill=onboardBand, rounded corners=6pt, fit=(o1)(o3),
          inner xsep=5mm, inner ysep=5mm] (bandOnboard) {};
    \node[fill=qcBand, rounded corners=6pt,
          fit=(assign_qc)(assign_file)(expected)(unexpected)(sample)(allfiles)(decision),
          inner xsep=5mm, inner ysep=5mm] (bandQC) {};
    \node[fill=acceptBand, rounded corners=6pt, fit=(accepted)(speechAcc),
          inner xsep=3mm, inner ysep=3mm] (bandAccept) {};
    \node[fill=rejectBand, rounded corners=6pt, fit=(rejected)(speechRej),
          inner xsep=3mm, inner ysep=3mm] (bandReject) {};
\end{scope}

% Stage tags
\node[stagetag, fill=onboardBorder, text=white, anchor=south west]
    at ([xshift=4mm, yshift=1mm]bandOnboard.north west){PHASE 1 \textbar\ Onboarding};
\node[stagetag, fill=qcBorder, text=white, anchor=south west]
    at ([xshift=4mm, yshift=1mm]bandQC.north west){PHASE 2 \textbar\ Quality-Control Workflow};
\node[stagetag, fill=acceptBorder, text=white, anchor=south west]
    at ([xshift=4mm, yshift=1mm]bandAccept.north west){\small ACCEPTED};
\node[stagetag, fill=rejectBorder, text=white, anchor=south west]
    at ([xshift=4mm, yshift=1mm]bandReject.north west){\small REJECTED};
\end{tikzpicture}%
}

\caption{Freelancer-based quality-control workflow for the VAANI speech corpus. Onboarded freelancers perform QC on assigned audio files; expected responses are reviewed on a 10\% sample basis, while unexpected responses trigger full review, with final accept/reject decisions made centrally.}
\label{fig:qc-workflow}
\end{figure}
\subsection{Manual Quality Control}
The files flagged during the automated check are selected for manual review. Additionally, we ensure that at least one segment from each speaker is included. These segments are randomly sampled so that 10\% of the data that passed both the Level 1 and Level 2 checks also undergoes manual quality control.

To ensure effective quality control (Figure \ref{fig:qc-workflow}), we have assembled and trained a team of  validators  across the districts, supervised by experienced data annotation professionals hailing from different  native languages. Data validation is carried out at the district level by validators from each respective district.  The first level of quality checks is conducted by the experts, with random samples being reviewed by a supervisor. If quality issues are identified greater than a set threshold (10 \%), the entire dataset is sent for further validation by additional experts to ensure accuracy and consistency. Manual quality checks are conducted independently for both the audio and the transcriptions. \\

In this manual quality check for audio, several important aspects are assessed to ensure clarity and relevance. First, it is confirmed that humans are speaking in the audio and that only one person is speaking about the image. The speaker's gender  is identified, and the language expert affirms that the audio is understandable. Additionally, the speaker is identified as someone from the district, ensuring contextual accuracy. The audio is free from disturbances that could hinder comprehension, and it sounds like a complete, coherent sentence. Furthermore, it is verified that the audio content is relevant to the image, and no flags were raised on the speaker's appropriateness. Privacy and confidentiality are also prioritized by ensuring there is no personally identifiable information (PII) in the image or audio files. This process helps maintain high standards of transcription and ensures the quality and relevance of the audio content.

Transcription quality is verified through a manual review conducted by a language expert. The reviewer confirms that the transcribed text matches the audio verbatim and that the script is consistent with the spoken language. They also check for word-level repetition and verify that the speech is natural rather than read or guided. Finally, they flag any personally identifiable information (PII) present in the audio or associated image. Together, these checks ensure transcription accuracy, fluency, and privacy compliance

\section{Experiments: Multilingual ASR Fine-Tuning}
 
To demonstrate the efficacy of VAANI, we fine-tune three open-source ASR models with distinct architectures: Gemma-3n-E2B~\cite{gemma_3n_2025}, Whisper-large-v3-turbo~\cite{radford2022robustspeechrecognitionlargescale}, and Parakeet-tdt-0.6b-v2~\cite{nvidia2025parakeettdt06bv2}. All models are trained in a multilingual setting on four well-represented languages, Hindi (593.93~Hrs), Bengali (113.24~Hrs), Kannada (120.20~Hrs), Telugu (113.04~Hrs) and two low-resource languages, Chakma (31.58~Hrs) and Bhojpuri (23.69~Hrs). The dataset is split into 80\% train, 10\% validation, and 10\% test within each district, ensuring no speaker overlap across splits.

\textbf{Gemma-3n-E2B} uses PEFT with 4-bit quantization and LoRA ($r{=}64$, $\alpha{=}128$) on both language and audio modules, trained with 8-bit AdamW (lr $5{\times}10^{-5}$, cosine schedule, 10\% warmup) for one epoch at effective batch size 128, with loss masked to textual outputs only. \textbf{Whisper-large-v3-turbo} is fully fine-tuned in a seq2seq setup with Adam (lr $1{\times}10^{-5}$), batch size 64, FP16, for up to 20 epochs with WER-based early stopping; for Indic languages unsupported by Whisper, inputs are mapped to linguistically related fallback languages via forced decoder tokens. \textbf{Parakeet-tdt-0.6b-v2} is fully fine-tuned via NVIDIA NeMo with a custom SentencePiece tokenizer  using Adam (lr $1{\times}10^{-4}$) for up to 50 epochs.  Because the original Parakeet model is English 
only and its vocabulary does not cover Indic scripts, the decoder is 
reinitialized to accommodate the new tokenizer and zero-shot Indic 
evaluation is not meaningful. All experiments were conducted on NVIDIA H100 (80~GB) GPUs. 
\begin{figure}[h]
\includegraphics[width=0.99\textwidth]{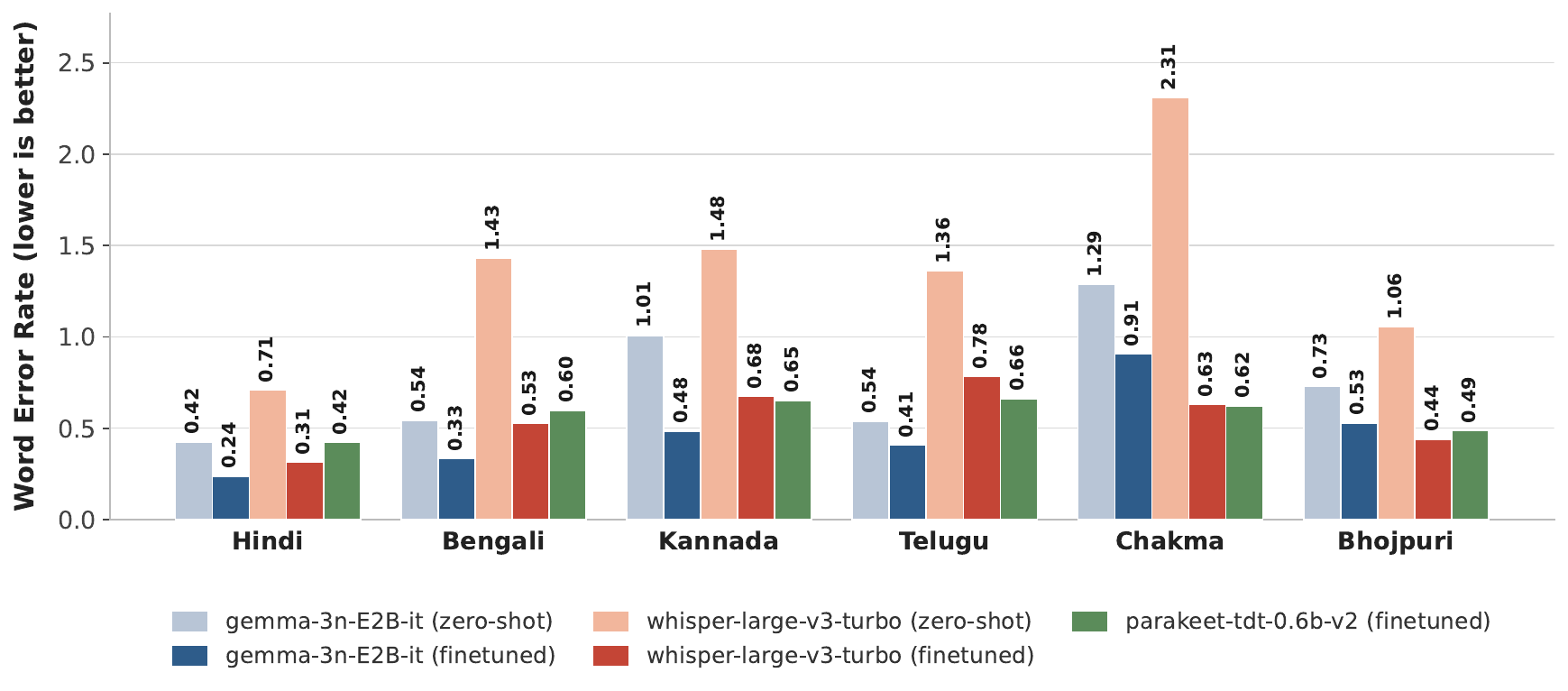}
\caption{Average WER of finetuned models on benchmark datasets across languages.}
\label{fig:experiment_finetuning}
\end{figure}
 
We evaluate on external benchmarks per language: Hindi (GramVAANI~\cite{bhanushali22_interspeech}, FLEURS, MUCS~\cite{Diwan_2021}, CommonVoice, Kathbath, Kathbath-Noisy~\cite{javed2023indicsuperb}, VAANI, RESPIN~\cite{kumarrespin}); Bengali and Telugu (FLEURS, CommonVoice, Kathbath, Kathbath-Noisy, VAANI); Kannada (FLEURS, Kathbath, Kathbath-Noisy, VAANI, RESPIN); and Chakma and Bhojpuri on VAANI only, as no public benchmarks exist. In the multilingual setting, models must infer the language and produce correctly-scripted transcripts without conditioning. Non-finetuned baselines frequently fail on under-resourced Indic languages, yielding incorrect scripts or repetitive hallucinations (Figure~\ref{fig:experiment_finetuning}). Fine-tuning on VAANI substantially improves model performance across both major and low-resource languages, demonstrating the dataset's utility across the resource spectrum.

\section{Applications and Limitations}
VAANI's scale, linguistic breadth, and multimodal structure support a range of downstream applications: ASR for low-resource Indic languages, end-to-end and cascaded speech translation, region-specific and culturally grounded benchmarking, voice conversion and speaker-adaptive TTS, and multimodal audio-visual modeling enabled by the aligned image–speech–text triplets. Several limitations remain, however. Audio and transcription coverage is small for many of the 105 languages, with the bulk of the data concentrated in a few majority languages. Geographic and linguistic coverage is also not yet exhaustive; VAANI spans 165 of roughly 800 districts, leaving many regional dialects and minority languages unrepresented. We also note the dual-use risk that speaker-diverse audio could be misused for voice cloning or impersonation, and apply mitigations including speaker ID anonymization, removal of personally identifiable information, and consent-based data collection.
\section{Conclusions and Future work}
In summary, we present VAANI, a rigorously curated multimodal dataset comprising 31,255 hours of audio, 2,043 hours of transcriptions, and 289,838 images, enriched with extensive metadata across 165 districts and 105 languages. Our preliminary experiments confirm the dataset's utility across speech and multimodal tasks, and its broad linguistic diversity enables the development of accent-diverse models that generalize beyond standard dialects while supporting ASR research for low-resource languages. 
By open-sourcing VAANI under a permissive license, we hope to lower the barrier to entry for Indic speech research and accelerate the development of voice technologies that meaningfully serve underrepresented linguistic communities across India. We view VAANI as a foundational resource for the broader research community and are actively working on future phases to improve audio and transcription coverage for low-resource languages and to expand the dataset to additional districts. 
%\input{sections/limitations}
%\bibliographystyle{plainnat}
%\bibliography{tacl2018}
%\bibliographystyle{plainnat}
\bibliographystyle{unsrt}
\bibliography{tacl2018}

\newpage
\appendix
\section{Experiments}
The VAANI dataset is a multimodal corpus comprising approximately 2,043 hours of transcribed speech across various Indian languages. To demonstrate its practical utility, we conducted preliminary experiments using a selected subset of its speech and visual modalities.

\subsection{ASR Finetuning : Language Specific}
To demonstrate the utility of the VAANI dataset, we evaluate the performance gains achieved by fine-tuning a pre-trained baseline model on a selected subset of the data. Specifically, we independently fine-tuned the Whisper-small model (244M parameters) \cite{radford2022robustspeechrecognitionlargescale} for four languages: Hindi (331 hours), Bengali (101 hours), Kannada (80.2 hours), and Telugu (69.0 hours). The resulting models were then evaluated across multiple benchmark datasets to assess the consistency of these improvements.

All experiments were conducted using NVIDIA L40S GPUs. Implementation was carried out using the Hugging Face Transformers library. During preprocessing, transcripts underwent text normalization and were capped at a maximum sequence length of 448 tokens. We approached training within a sequence-to-sequence framework, utilizing the Adam optimizer with a learning rate of $1 \times 10^{-5}$  and 1,000 warmup steps. To optimize computational efficiency, we employed mixed-precision (FP16) training for 10 epochs. A per-device batch size of 16 was used alongside a gradient accumulation step of 2, yielding an effective batch size of 32.

Experimental results demonstrate consistent performance improvements across all four languages compared to the pre-trained open-source baseline, thereby illustrating the VAANI dataset's efficacy for fine-tuning robust ASR models.
\begin{figure*}[htbp]
    \centering
    \begin{subfigure}[b]{0.45\textwidth}
        \includegraphics[width=\textwidth]{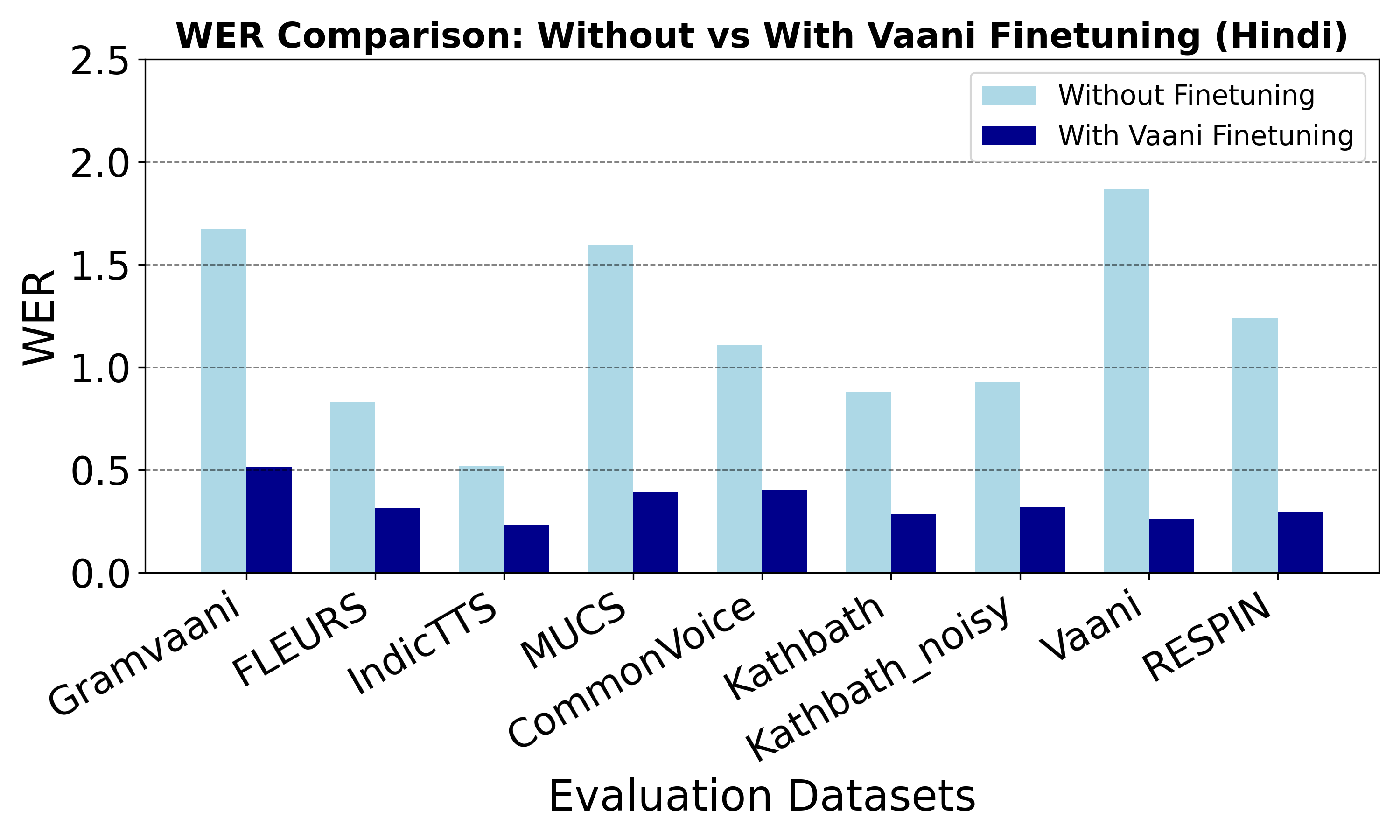}
        \caption{Hindi}
    \end{subfigure}
    \hfill
    \begin{subfigure}[b]{0.45\textwidth}
        \includegraphics[width=\textwidth]{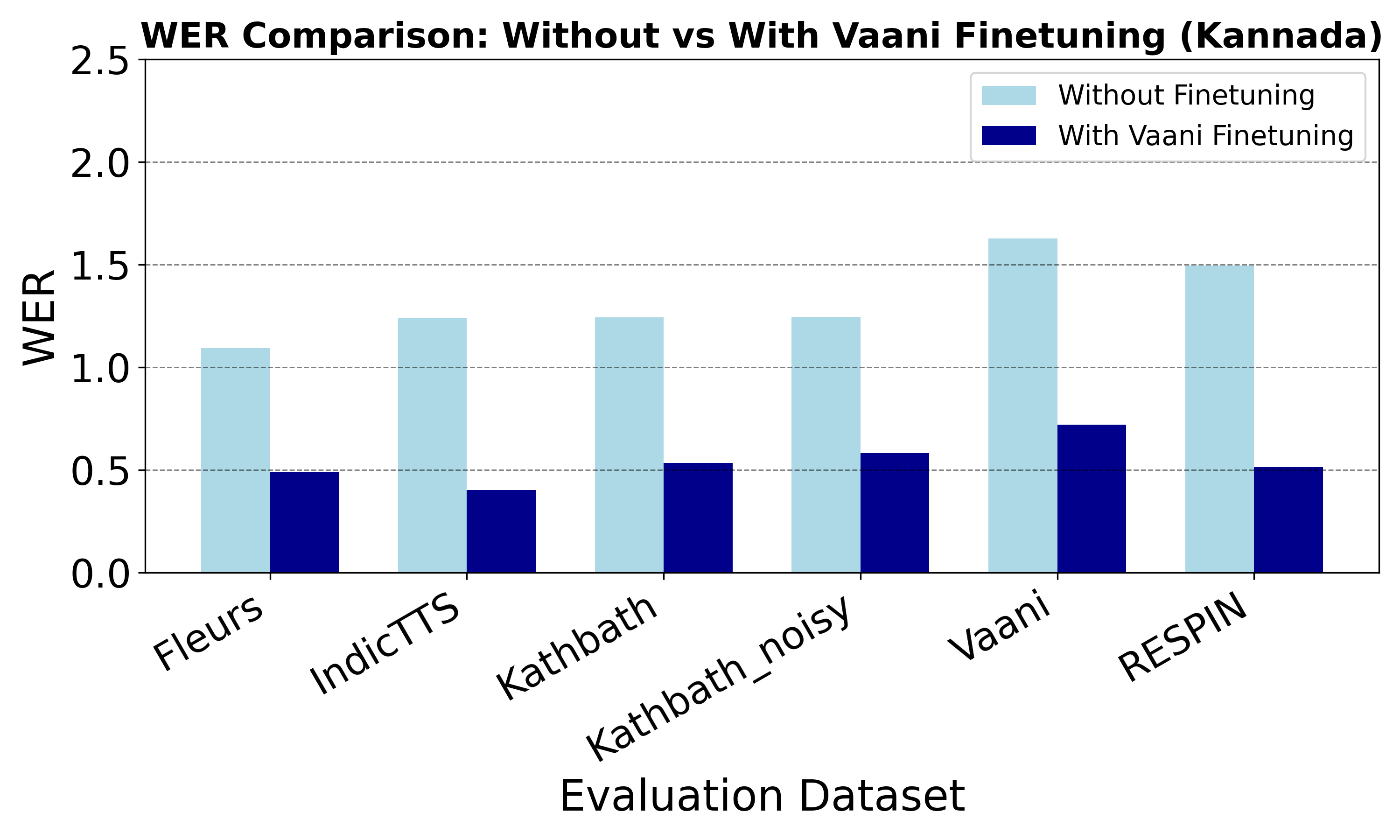}
        \caption{Kannada}
    \end{subfigure}

    \vspace{0.5em} % optional vertical spacing

    \begin{subfigure}[b]{0.45\textwidth}
        \includegraphics[width=\textwidth]{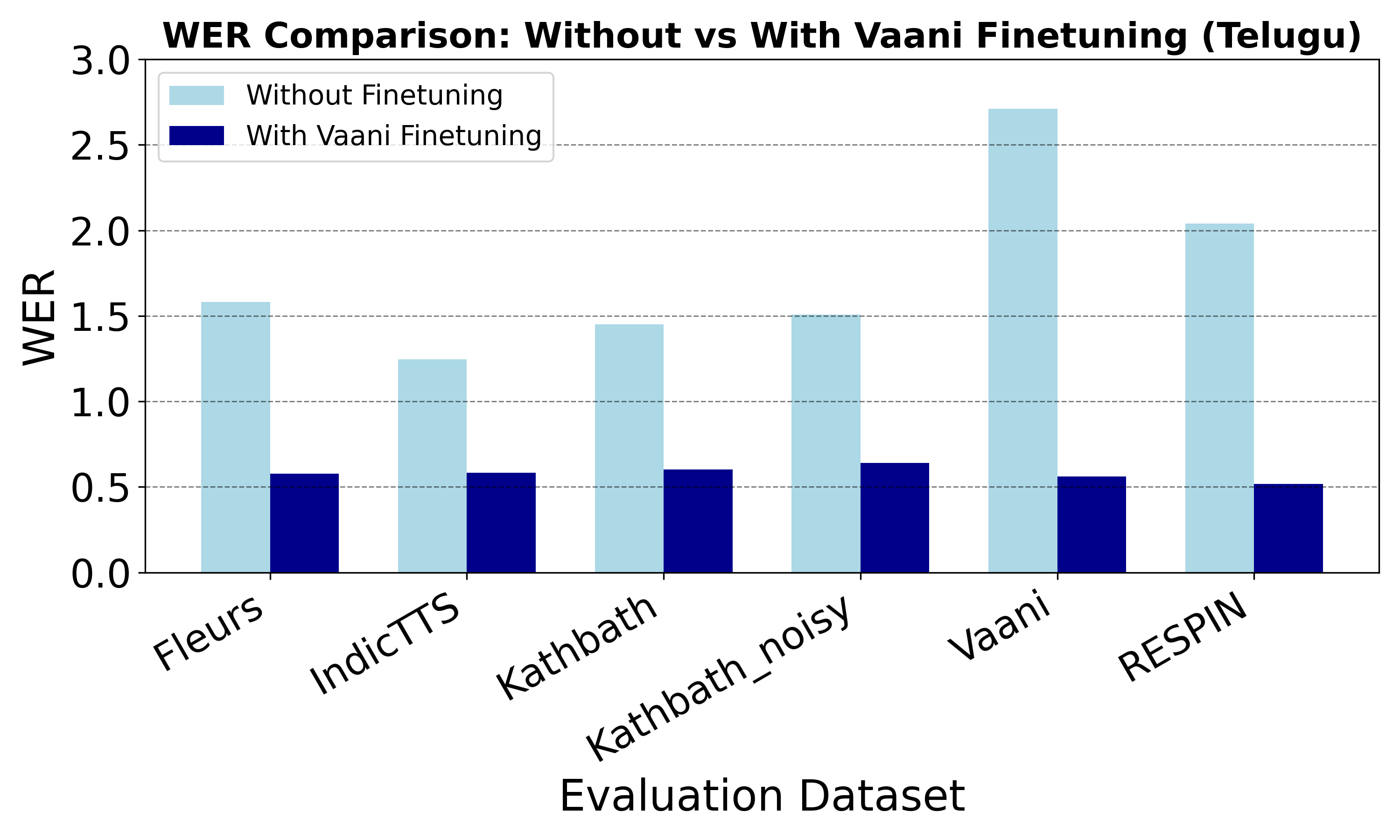}
        \caption{Telugu}
    \end{subfigure}
    \hfill
    \begin{subfigure}[b]{0.45\textwidth}
        \includegraphics[width=\textwidth]{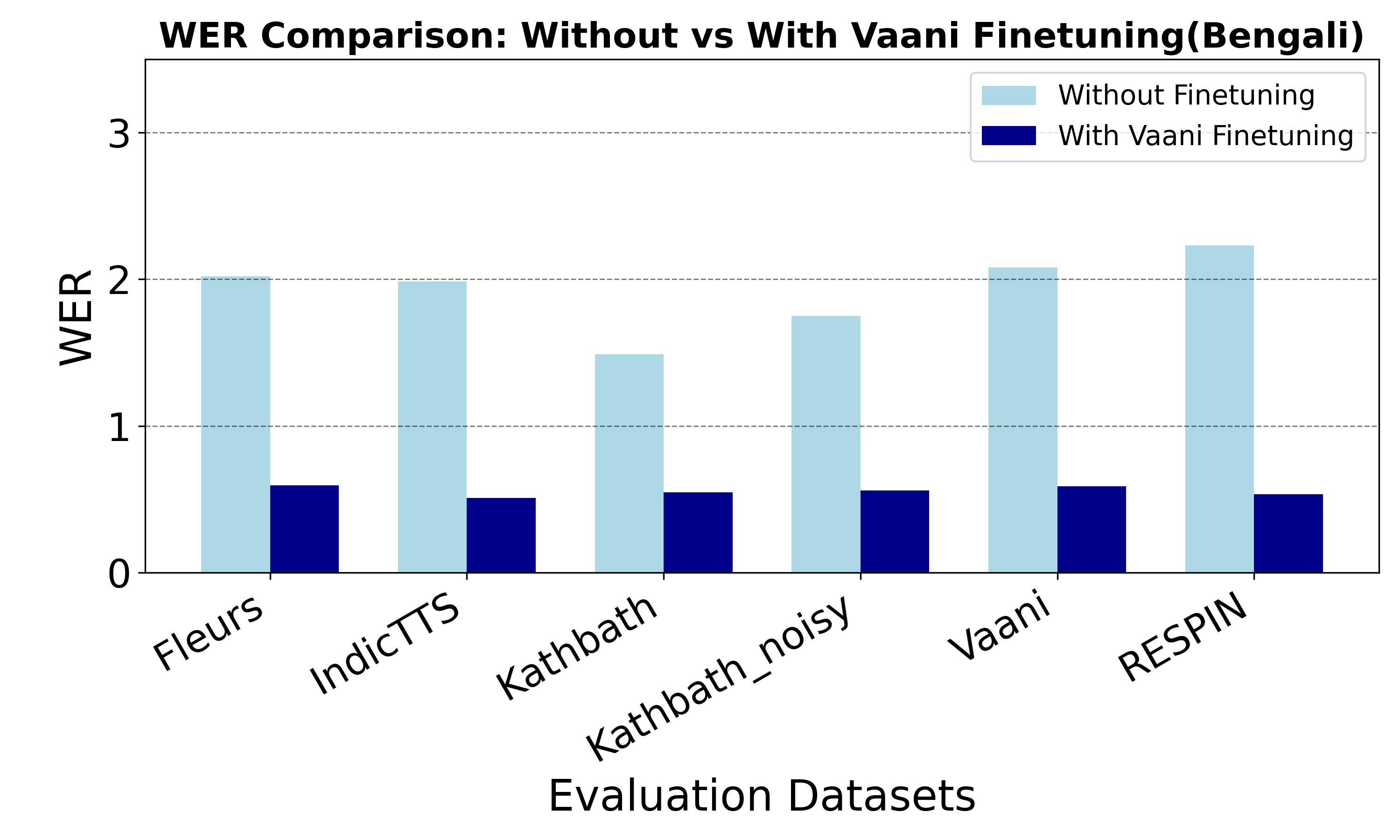}
        \caption{Bengali}
    \end{subfigure}

    \caption{Word Error Rate (WER) comparison for Whisper-small models fine-tuned on the VAANI dataset across multiple languages. Each subplot shows performance on benchmark datasets for the respective language.}
    \label{fig:wer_finetuning}
\end{figure*}

\subsection{Region Specific finetuning}
In order to highlight the importance of region-specific data, we fine-tuned models for the Hindi dataset at the state level, and the performance was measured across multiple states.
\begin{figure*}
\includegraphics[width=0.95\textwidth]{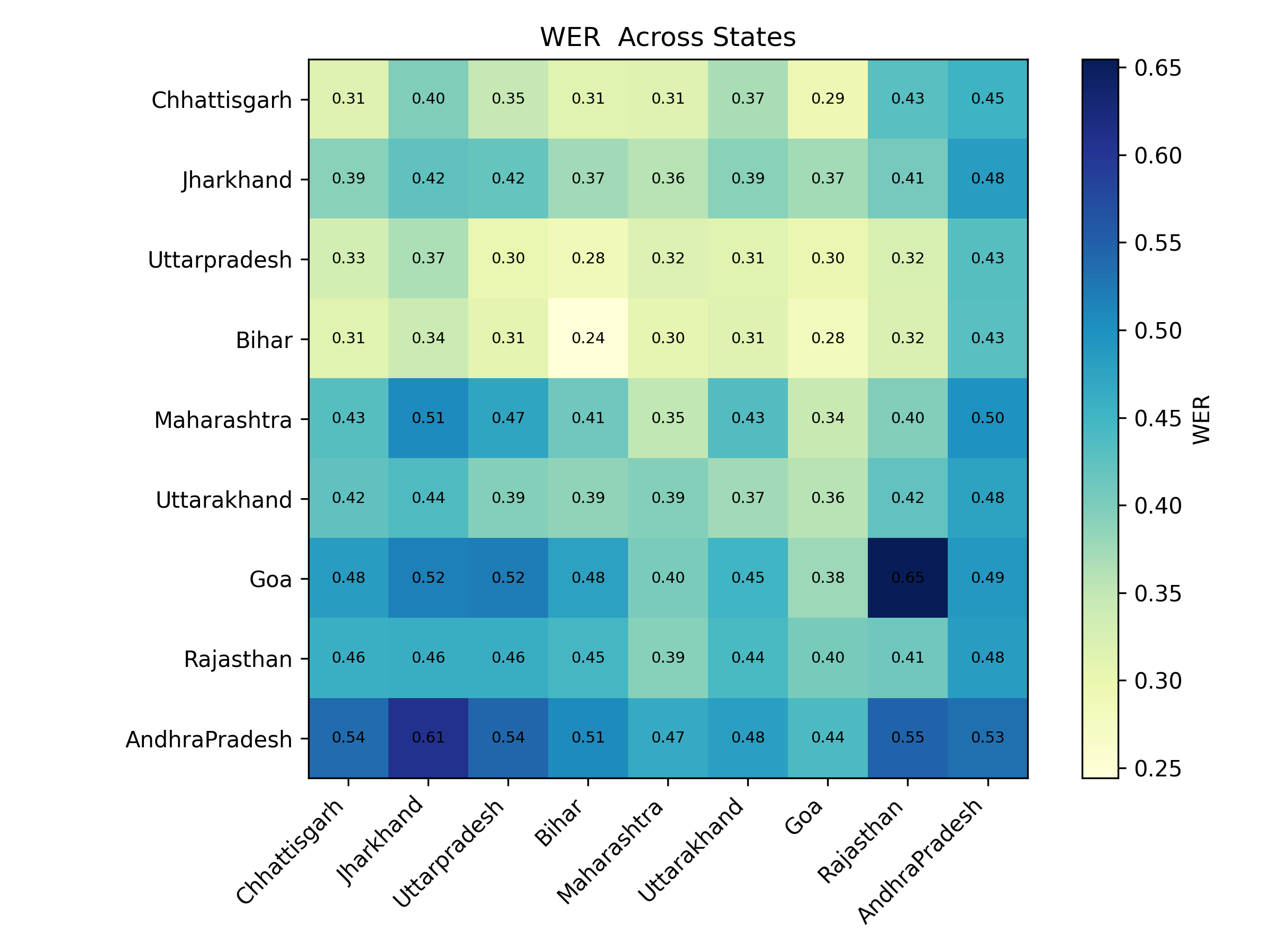}
\caption{WER heatmap showing state-wise evaluation results for Hindi models fine-tuned on data from individual states. The results highlight that models tend to perform better on geographically proximate states compared to distant ones, even within the same language.}
% \label{fig:lid_cm1}
\end{figure*}
The model is fine-tuned using the Whisper small architecture within the Hugging Face Transformers framework for automatic speech recognition.  Training is performed using a sequence-to-sequence framework with a per-device batch size of 16 and gradient accumulation of 2, resulting in an effective batch size of 32. The model is optimized using the Adam optimizer with a learning rate of $1 \times 10^{-5}$ and a warmup of 1000 steps. Training is conducted for 10 epochs with mixed-precision (FP16) to improve computational efficiency. Evaluation is performed at regular intervals of 200 steps using word error rate (WER) as the primary metric, and the best model is selected based on minimum WER. The maximum generation length is set to 225 tokens during decoding. Model checkpoints are saved every 200 steps with a limit of four checkpoints, and training progress is monitored.

We fine-tuned the model on Hindi data separately for each state—Bihar (130.02 hours), Uttar Pradesh (75.63 hours), Chhattisgarh (60.45 hours), Maharashtra (21.81 hours), Jharkhand (15.63 hours), Uttarakhand (10.59 hours), Rajasthan (5.54 hours), Goa (4.88 hours), and Andhra Pradesh (4.49 hours)—and evaluated performance using test data from the same states.

The evaluation results show that a model trained on data from a specific state performs comparatively better on nearby states than on distant ones, even within the same language.

\subsection{Image Retrieval }

We aimed to explore the image modality within the VAANI dataset, which contains over 289K  images collected specifically for this purpose. Since speakers were prompted to speak about images, the corresponding audio often includes details related to the visual content. Multiple speakers may describe the same image, and parts of these utterances are transcribed. For this experiment, we aggregated the transcribed segments for each image and used Large language Model(Gemini 2.5 -Flash) to generate a single consolidated description.
\begin{figure}[H]
\centering
\includegraphics[width=1.0\textwidth]{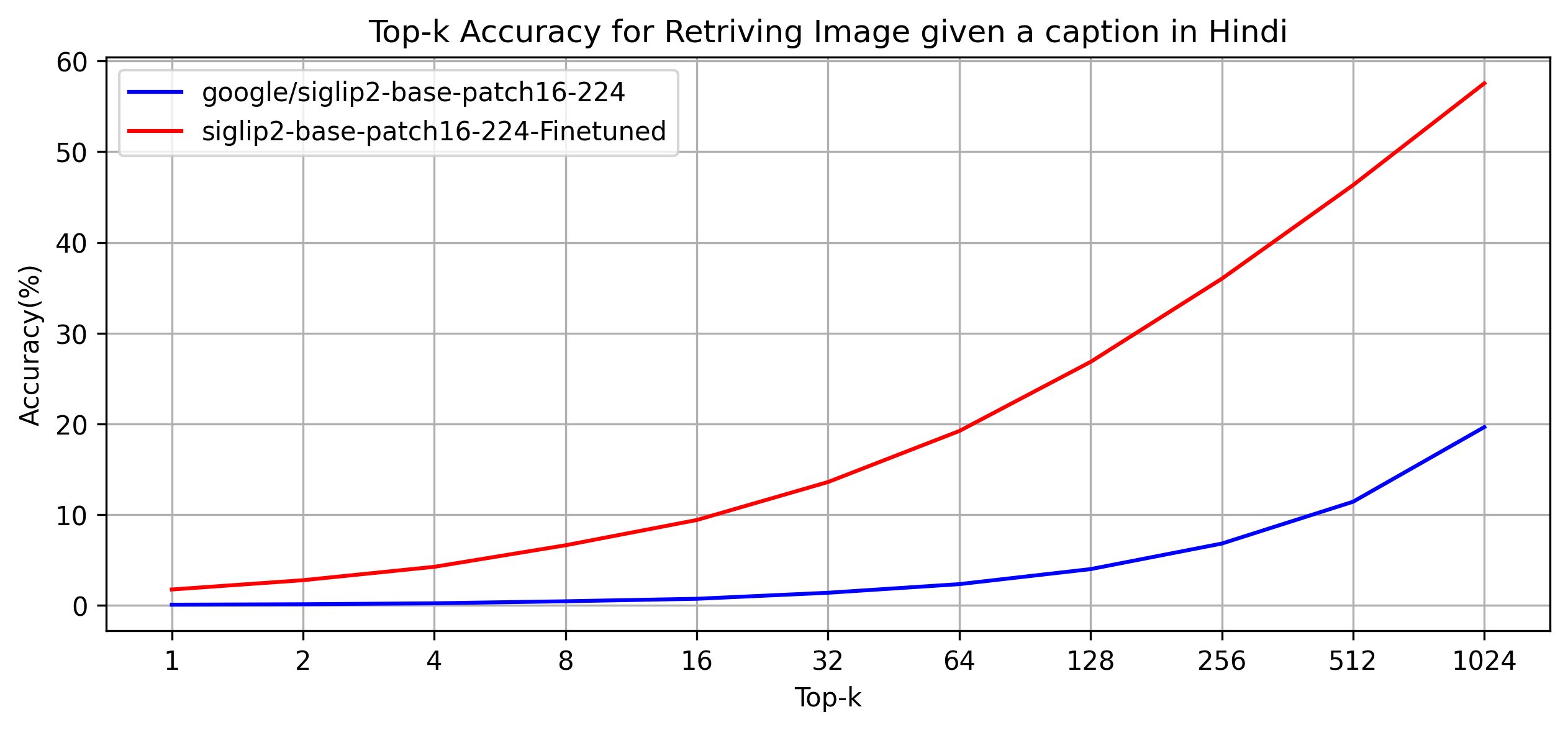} 
\caption{Performance of SigLIP2 fine-tuned on the Hindi portion of the VAANI dataset for the image retrieval task.}
\label{fig:siglip2_perf}
\end{figure}
For the experiments, we used the SigLIP2 \cite{tschannen2025siglip} model and the Hindi portion of the dataset. The model was fine-tuned on image–Hindi transcription pairs, using approximately 45,000 images for training, and evaluated on an image retrieval task with around 9,500 test images. The fine-tuned model demonstrated performance improvements over the original model.

% Required packages (add to preamble if not already present):
%   \usepackage{graphicx}
%   \usepackage{subcaption}
%   \usepackage{tikz}
%   \usetikzlibrary{arrows.meta, positioning, calc}
%   \usepackage{pifont}        % for \ding
%   \usepackage{xcolor}
%   \usepackage{amsmath}

\section{Image Data Specifications and Collection Guidelines}
In order to collect and curate image data that facilitates the speech data collection process, we define a detailed set of specifications governing how images are captured, filtered, and delivered by external vendors responsible for data acquisition. In this setup, the vendor is tasked with sourcing and curating images in accordance with these guidelines, ensuring quality control, diversity, and regional relevance. The Vaani dataset uses these images as visual prompts: a speaker is shown an image and asked to describe it in their own words, thereby eliciting natural, spontaneous speech grounded in real-world visual content. Consequently, the quality, diversity, and regional authenticity of vendor-collected images directly influence the richness and representativeness of the resulting speech corpus.
\begin{figure}[htbp]
    \centering
    % ========== PART 1: 2x2 Image Grid ==========
    \begin{subfigure}[b]{0.48\textwidth}
        \centering
        \begin{tabular}{@{}cc@{}}
            \includegraphics[width=0.48\linewidth]{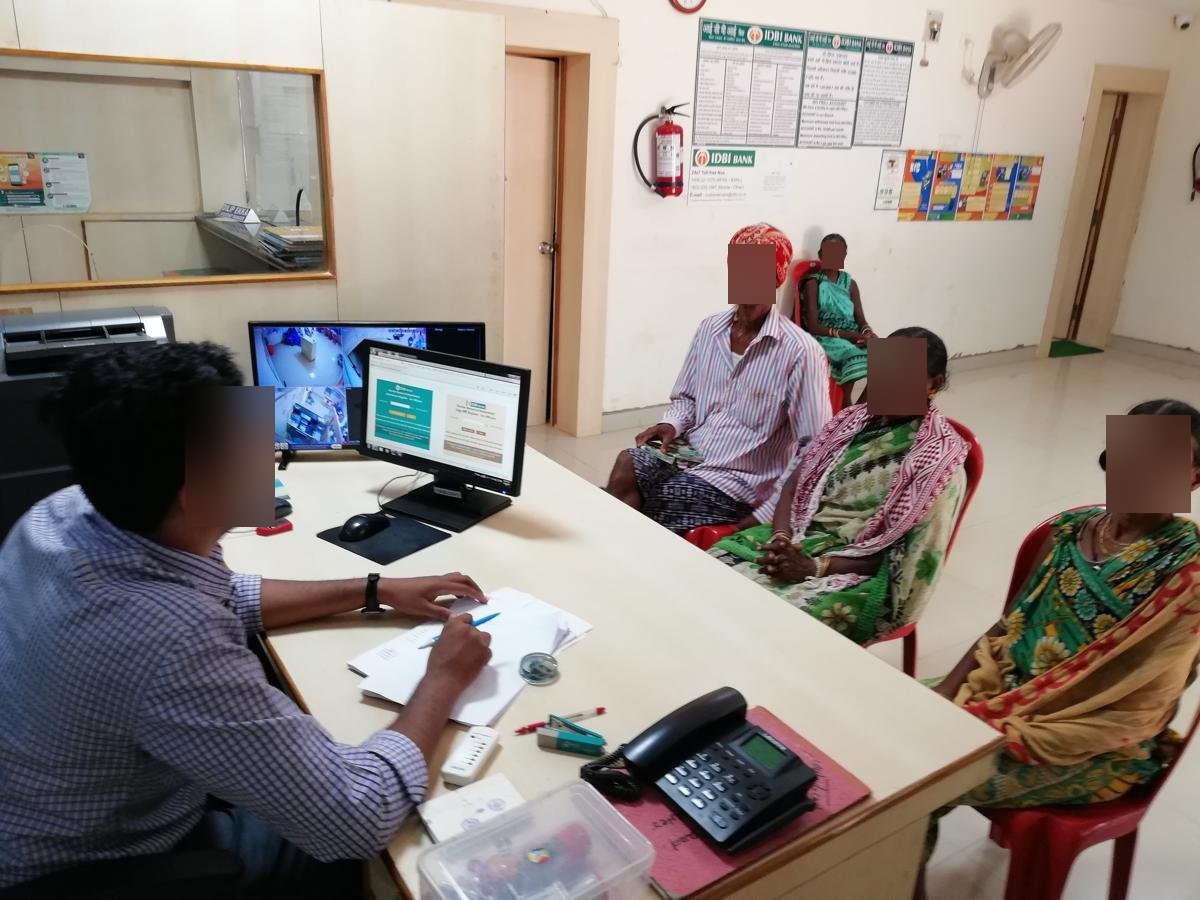} &
            \includegraphics[width=0.48\linewidth]{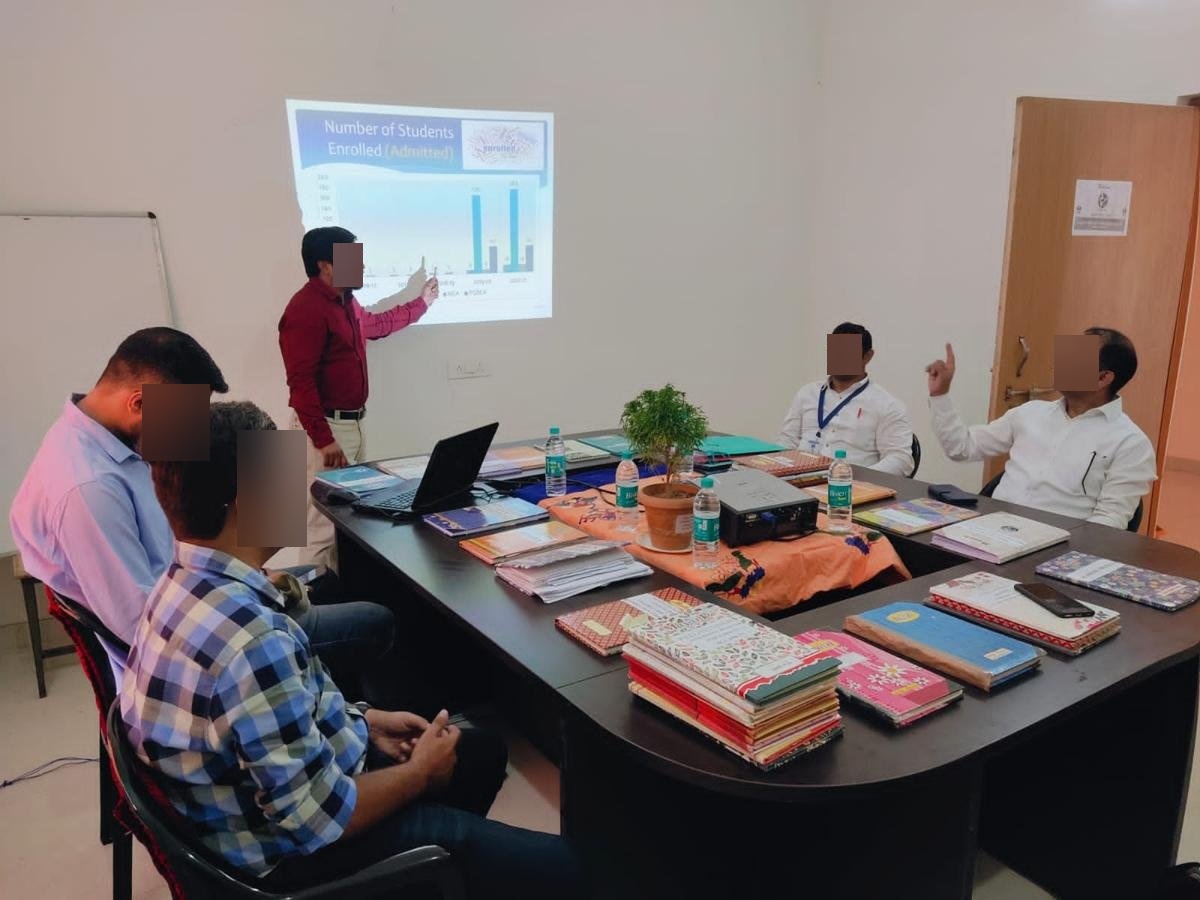} \\[2pt]
            \includegraphics[width=0.48\linewidth]{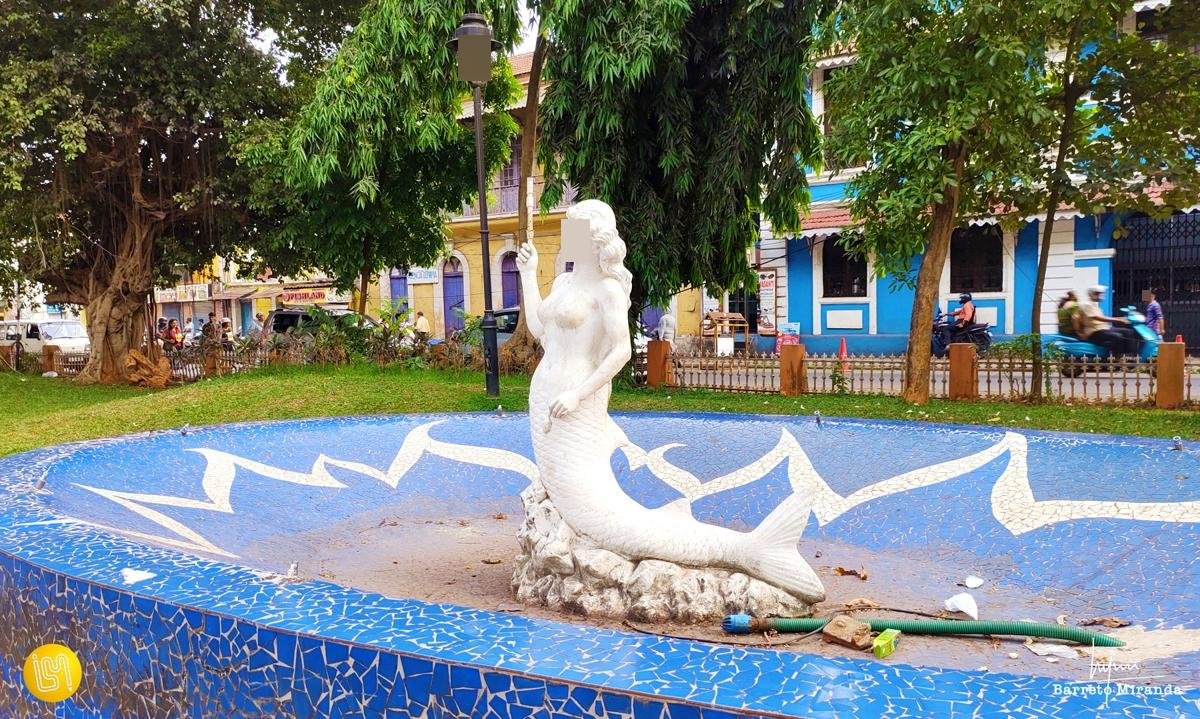} &
            \includegraphics[width=0.48\linewidth]{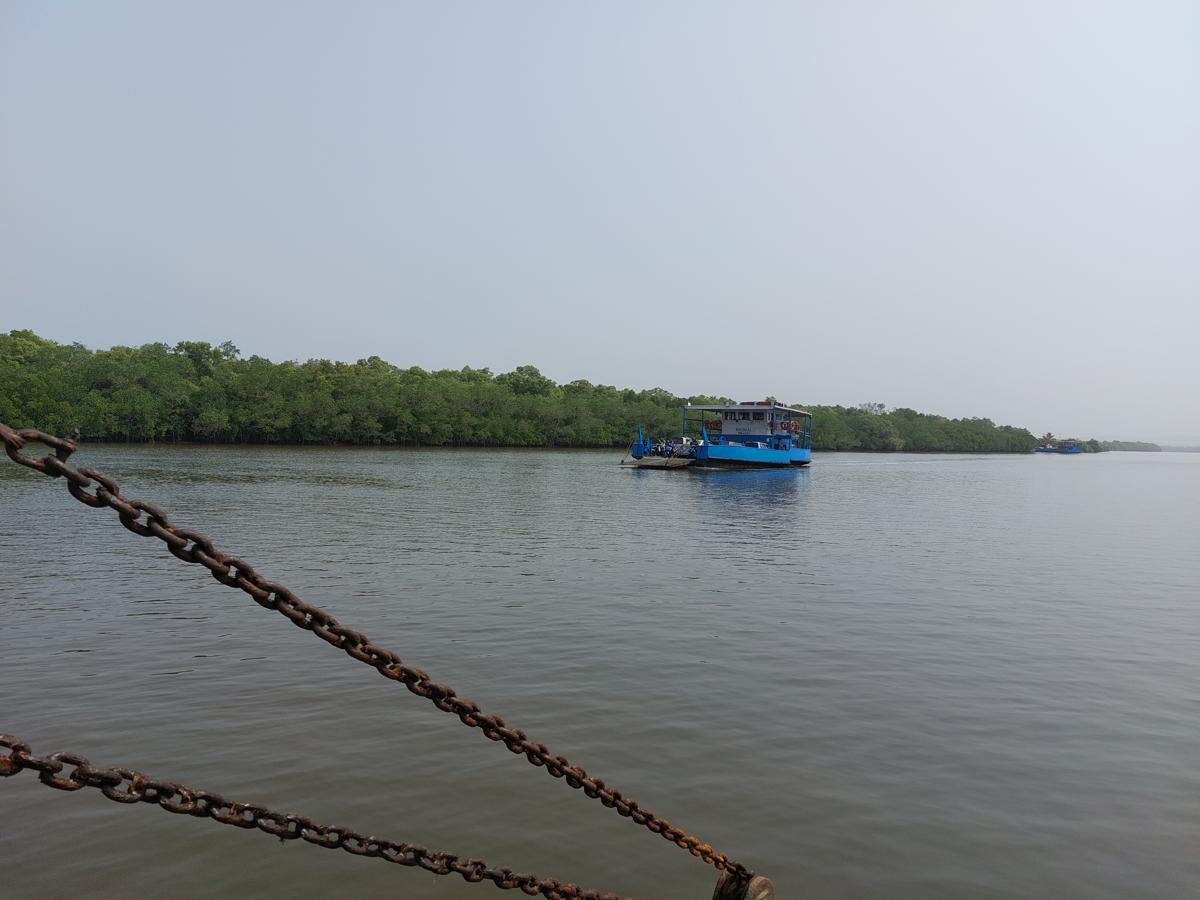} \\
        \end{tabular}
        \caption{Sample images captured for a district.}
        \label{fig:sample_images}
    \end{subfigure}
    \hfill
    % ========== PART 2: Block Diagram ==========
    \begin{subfigure}[b]{0.48\textwidth}
        \centering
        \vspace{0pt}
        \begin{tikzpicture}[
            node distance=0.6cm and 0.4cm,
            block/.style={
                rectangle, rounded corners=4pt, draw=blue!60!black, thick,
                fill=blue!10, minimum width=3.2cm, minimum height=0.9cm,
                align=center, font=\small\bfseries
            },
            finalblock/.style={
                rectangle, rounded corners=4pt, draw=green!50!black, thick,
                fill=green!15, minimum width=3.2cm, minimum height=0.9cm,
                align=center, font=\small\bfseries
            },
            arrow/.style={-{Stealth[length=2.5mm]}, thick, blue!60!black}
        ]
            \node[block] (capture) {Image Capture};
            \node[block, below=of capture] (quality) {Quality Check};
            \node[block, below=of quality] (content) {Content Check};
            \node[finalblock, below=of content] (selected) {Selected Images};

            \draw[arrow] (capture) -- (quality);
            \draw[arrow] (quality) -- (content);
            \draw[arrow] (content) -- (selected);
        \end{tikzpicture}
        \caption{Image curation pipeline.}
        \label{fig:pipeline}
    \end{subfigure}

    \caption{Data collection overview: (a) representative images captured in the field, and (b) the pipeline applied to filter captured images down to the final selected set.}
    \label{fig:data_collection}
\end{figure}

\subsection{Type of Images and Geography}
\begin{itemize}
    \item \textbf{Specific Images:} Images must be physically shot from the target district. These images must relate to a site specific to that district, rather than depicting a generic topic.
\end{itemize}

\subsection{Image Topics and Distribution}
\begin{itemize}
    \item \textbf{Distribution:} All topics provided must contain an equal number of images, allowing for a tolerance of $\pm 20\%$ in quantity variation.
\end{itemize}

\subsection{Image Specifications}
\begin{itemize}
    \item \textbf{Format:} \texttt{.jpg}.
    \item \textbf{Size and Aspect Ratio:} $640 \times 400$ pixels or a 16:9 aspect ratio.
    \item \textbf{File Size:} $< 500$~KB.
    \item \textbf{Quantity:} The number of images per topic per district must align with the image topic table provided.
    \item \textbf{Metadata:} The date of shooting must be embedded in the image file properties and must be no earlier than July 1, 2023.
\end{itemize}

\subsection{Collection Guidelines and Quality Control}
\begin{itemize}
    \item \textbf{Authenticity:} Images must be shot physically in the real world. Capturing an image of an existing image is strictly prohibited. Sourcing from secondary or tertiary platforms (e.g., webpages, search engines, social media) is not allowed.
    \item \textbf{Recency:} The date of the image capture must be no earlier than July 1, 2023.
    \item \textbf{Clarity and Focus:} Images must be clear, well-focused (non-blurry), and free of any physical obstructions to the primary object. The object determined by the image category must be clearly visible.
    \item \textbf{Uniqueness:} Each image must be unique to a specific district and category; no repetitions are allowed across categories or districts. Additionally, file names must remain entirely unique, even across different delivery batches.
    \item \textbf{Privacy and PII:} The images must absolutely not contain any Personally Identifiable Information (PII). This includes, but is not limited to, faces of individuals, personally owned objects or properties, recognizable logos, or any object showcasing PII.
    \item \textbf{Describable Content:} The visual content must be rich and explicitly describable. Because the dataset's objective is to prompt a speaker to look at the image and talk about it, the image must contain sufficient visual elements to stimulate descriptive speech.
    \item \textbf{Quality Assurance (QA):} The vendor is required to establish strict internal quality checks and processes to adhere to these specifications. Deliveries may only proceed after internal verification is complete.
    \item \textbf{Batch Size:} The overall delivery batch size should be a minimum of 5{,}000 images and must not exceed 25{,}000 images. Note that the preferred batch size may be adjusted as the project progresses.
\end{itemize}

% ============================================================
% QC Pipeline Diagram
% ============================================================

% Custom colors for the QC pipeline diagram
\definecolor{blueFill}{RGB}{220, 235, 252}
\definecolor{blueBorder}{RGB}{74, 144, 226}
\definecolor{orangeFill}{RGB}{255, 232, 200}
\definecolor{orangeBorder}{RGB}{245, 166, 35}
\definecolor{purpleFill}{RGB}{232, 218, 239}
\definecolor{purpleBorder}{RGB}{155, 81, 224}
\definecolor{greenFill}{RGB}{215, 236, 209}
\definecolor{greenBorder}{RGB}{122, 184, 88}
\definecolor{redFill}{RGB}{251, 215, 215}
\definecolor{redBorder}{RGB}{217, 83, 79}
\definecolor{coralFill}{RGB}{255, 200, 180}
\definecolor{coralBorder}{RGB}{240, 130, 90}
\definecolor{acceptedGreen}{RGB}{60, 150, 60}
\definecolor{rejectedRed}{RGB}{200, 50, 50}

\begin{figure*}[htbp]
    \centering
    \resizebox{\textwidth}{!}{%
    \begin{tikzpicture}[
        node distance=0.7cm and 0.5cm,
        font=\sffamily\footnotesize,
        block/.style={
            rectangle, rounded corners=3pt, draw, thick,
            minimum width=1.9cm, minimum height=1.0cm,
            align=center, text width=1.9cm, inner sep=3pt
        },
        blueblock/.style={block, fill=blueFill, draw=blueBorder},
        orangeblock/.style={block, fill=orangeFill, draw=orangeBorder},
        purpleblock/.style={block, fill=purpleFill, draw=purpleBorder},
        greenblock/.style={block, fill=greenFill, draw=greenBorder},
        redblock/.style={block, fill=redFill, draw=redBorder},
        coralblock/.style={block, fill=coralFill, draw=coralBorder},
        arrowblack/.style={-{Stealth[length=3mm]}, thick, black!70},
        arrowgreen/.style={-{Stealth[length=3mm]}, thick, acceptedGreen},
        arrowred/.style={-{Stealth[length=3mm]}, thick, rejectedRed},
        acceptedlabel/.style={text=acceptedGreen, font=\sffamily\scriptsize, fill=white, inner sep=1pt},
        rejectedlabel/.style={text=rejectedRed, font=\sffamily\scriptsize, fill=white, inner sep=1pt}
    ]

    % Top row of main pipeline
    \node[blueblock] (vendor) {Vendor};
    \node[blueblock, right=of vendor] (batch) {Batch received\\from Vendor};
    \node[orangeblock, right=of batch] (initial) {Initial Checks\\Filename, Size,\\Extension, Dimensions};
    \node[purpleblock, right=1.4cm of initial] (automated) {Automated Checks\\Blur, Edge, Histogram,\\Face, Human Detection};
    \node[greenblock, right=1.4cm of automated] (final) {Run Final Selection};
    \node[blueblock, right=of final] (upload) {Upload Accepted\\Images to Bucket};

    % Rejected node from Initial Checks (below)
    \node[redblock, below=1.2cm of initial] (rej1) {\textcolor{rejectedRed}{\textbf{\ding{55}}} Rejected\\(Initial Checks)};

    % Manual QC chain (below Automated Checks)
    \node[orangeblock, below=1.2cm of automated] (createlist) {Create List\\for Manual QC};
    \node[coralblock, below=1.0cm of createlist] (manualqc) {Manual QC\\Human Review\\Privacy \& Quality};

    % Rejected node from Manual QC
    \node[redblock] (rej2) at ($(manualqc.east)+(4.6,-0.0)$) {\textcolor{rejectedRed}{\textbf{\ding{55}}} Rejected\\(Manual QC)};

    % Top horizontal arrows
    \draw[arrowblack] (vendor) -- (batch);
    \draw[arrowblack] (batch) -- (initial);

    % Initial Checks -> Automated Checks (Accepted)
    \draw[arrowgreen] (initial) -- (automated)
        node[acceptedlabel, midway, above] {Accepted};

    % Automated Checks -> Final Selection (Accepted)
    \draw[arrowgreen] (automated) -- (final)
        node[acceptedlabel, midway, above] {Accepted};

    % Final Selection -> Upload
    \draw[arrowblack] (final) -- (upload);

    % Initial Checks -> Rejected (down)
    \draw[arrowred] (initial) -- (rej1)
        node[rejectedlabel, midway, right] {Rejected};

    % Automated Checks -> Create List (Rejected, down)
    \draw[arrowred] (automated) -- (createlist)
        node[rejectedlabel, midway, right] {Rejected};

    % Create List -> Manual QC
    \draw[arrowblack] (createlist) -- (manualqc);

    % Manual QC -> Final Selection (Accepted)
    \coordinate (turnUp) at ($(manualqc.east)+(0.6,0)$);
    \draw[arrowgreen] (manualqc.east) -- (turnUp) -- (turnUp |- final.west) -- (final.west);
    \node[acceptedlabel] at ($(turnUp)+(0,1.2)$) {Accepted};

    \draw[arrowred] (manualqc) -- (rej2)
        node[rejectedlabel, midway, above] {Rejected};
    \end{tikzpicture}
    }%
    \caption{End-to-end image quality control pipeline. Batches received from the vendor pass through initial checks (filename, size, extension, dimensions) and automated checks (blur, edge, histogram, face, human detection). Items failing automated checks are routed to manual QC, after which accepted images are merged with the automated-accept set and uploaded to the bucket.}
    \label{fig:qc_pipeline}
\end{figure*}
\section{Audio Data Specifications and Collection Guidelines}

Audio data is collected with the support of vendors, who follow predefined specifications and detailed guidelines to ensure consistency, quality, and adherence to the data collection protocol.
%==============================================================
\subsection{Audio Specification}

\subsubsection{Format}
\begin{itemize}
    \item[1.1.1] 16\,kHz, 16 bits per sample
    \item[1.1.2] Single channel
    \item[1.1.3] Raw audio without any transcoding or post-processing
\end{itemize}

\subsubsection{Style}
\begin{itemize}
    \item[1.2.1] Spontaneous speech
    \item[1.2.2] Utterances spoken on an image given
    \item[1.2.3] Instruction should be given to ensure variety in the spoken utterance (to increase vocabulary).
    \item[1.2.4] Images should be randomly chosen from the pool of images given and be presented to the speaker for recording one by one.
    \item[1.2.5] For each image prompt recorded by a participant, an effective 10--20\,sec of speech data should be recorded.
\end{itemize}

\subsubsection{Image Corpus}
\begin{itemize}
    \item[1.3.1] A pool of images will be shared; some of these will be specific to the district and some will be common (across districts) and generic.
    \item[1.3.2] The images are an IP(Applicable while in the collection phase) and shall not be permitted to be used by the company unless explicit written approval.
    \item[1.3.3] The image names must be saved as-is without any changes at the time of delivery.
\end{itemize}

\subsubsection{Recording Background}
\begin{itemize}
    \item[1.4.1] A quiet environment (should not have TV sound, kitchen sound, children crying, water flowing, or a group of people right next to the recording, etc., in case of in-door recording; similarly, no high-intensity sound such as bird sound or vehicle sound for out-door recording).
    \item[1.4.2] If done in-door, the room MUST NOT be echoey.
    \item[1.4.3] Recording should not be done when the speaker is walking or running.
    \item[1.4.4] If the recording is done through a smartphone, vibration sound or any other notification sound should be disabled before the start of the recording.
    \item[1.4.5] Recording should not have unwanted distortions including, but not limited to, clipping, far-field effect, modulation due to DC-trend, zero segments / loss-of-signal issues, and unnecessary non-speech sounds.
\end{itemize}

\subsubsection{Microphone Placement}
\begin{itemize}
    \item[1.5.1] The recording device should be kept no farther than 2 feet from the mouth of the speaker.
    \item[1.5.2] The speaker mouth-to-recording-device distance and orientation should be as constant as possible during recording (good to keep the recording device at a fixed location).
    \item[1.5.3] Microphone should be in front of the speaker.
    \item[1.5.4] Microphone should be at least 1 foot away from the speaker.
\end{itemize}

\subsubsection{Speaker Specification}
\begin{itemize}
    \item[1.6.1] Uniform selection in the age range: 20--70 years.
    \item[1.6.2] Gender balanced in each district.
    \item[1.6.3] At least 800--820 speakers in each district.
    \item[1.6.4] A speaker should not contribute more than 15 minutes of effective speech.
\end{itemize}

\subsubsection{Speaker Selection Criteria}
\begin{itemize}
    \item[1.7.1] Should have an age between 20 and 70 years.
    \item[1.7.2] Must be a native of the pincode as reported in the metadata.
    \item[1.7.3] Should be encouraged to speak in the language/dialect used at their home with family members.
    \item[1.7.4] That the speaker is native of the location should be verified based on PAN card and/or Aadhaar card.
    \item[1.7.5] For every speaker, details of where they stayed and for how many years from birth should be noted and provided together with the recording.
    \item[1.7.6] At the start of each milestone phase, We will specify the sub-district level speaker sampling criteria.
\end{itemize}

\subsubsection{Metadata (for every speaker)}
\begin{itemize}
    \item[1.8.1] \textbf{Speaker\_ID}: Speaker ID (anonymized, e.g., \texttt{Spkr-A}, \texttt{Spkr-B}; unique to each speaker). Format example: \fmt{Speaker\_ID: speaker123\textbackslash n}. Allowed characters: \fmt{[a-z, A-Z, 0-9]}.
    \item[1.8.2] \textbf{Phone Brand and Phone Model}: Device ID (to capture device characteristics, e.g., phone brand and phone model). Format example: \fmt{Phone Brand: phoneX\textbackslash n}. Allowed characters: \fmt{[a-z, A-Z, 0-9]}.
    \item[1.8.3] \textbf{Gender}: Gender of the person. Format example: \fmt{Gender : Male\textbackslash n}. Allowed characters: \fmt{[a-z, A-Z]}.
    \item[1.8.4] \textbf{Age}: Age in years. Format example: \fmt{Age : 43\textbackslash n}. Allowed characters: \fmt{[0-9]}.
    \item[1.8.5] \textbf{Education}: Educational qualification (below 10th, 10th pass, 12th pass, graduate, postgraduate). Format example: \fmt{Education : 12th pass\textbackslash n}. Allowed characters: \fmt{[a-z, A-Z, 0-9]}.
    \item[1.8.6] \textbf{Speaker\_Pincode}: Pincode of the permanent address (where the person spent most of their life). Format example: \fmt{Speaker\_Pincode : 230023\textbackslash n}. Allowed characters: \fmt{[0-9]}.
    \item[1.8.7] \textbf{Socio\_economic}: Socio-economic status (Upper, Upper Middle, Lower Middle, Upper Lower, Lower; ref. \href{https://www.ncbi.nlm.nih.gov/pmc/articles/PMC6618222/}{NCBI PMC6618222}). Format example: \fmt{Socio\_economic : Upper Middle\textbackslash n}. Allowed characters: \fmt{[a-z, A-Z]}.
    \item[1.8.8] \textbf{Language\_during\_recording}: The name of the language/dialect the speaker speaks for the recording. Format example: \fmt{Language\_during\_recording : Marathi\textbackslash n}. Allowed characters: \fmt{[a-z, A-Z]}.
    \item[1.8.9] \textbf{Languages\_spoken}: Languages he/she can speak in general. This must also include the language in which the audio was recorded. Format example: \fmt{Languages\_spoken : Marathi,Hindi,English\textbackslash n}. Allowed characters: \fmt{[a-z, A-Z]}.
    \item[1.8.10] \textbf{Stay\_Years}: Details of where they stayed for how many years from birth. Format example: \fmt{Stay\_Years : NewDelhi(25),Mumbai(25)\textbackslash n}. Allowed characters: \fmt{[a-z, A-Z, 0-9]}.
\end{itemize}
\noindent\textit{Note: All fields must contain a newline character at the end.}

\begin{figure}[h]
    \centering
    \begin{subfigure}{0.23\textwidth}
        \includegraphics[width=\textwidth]{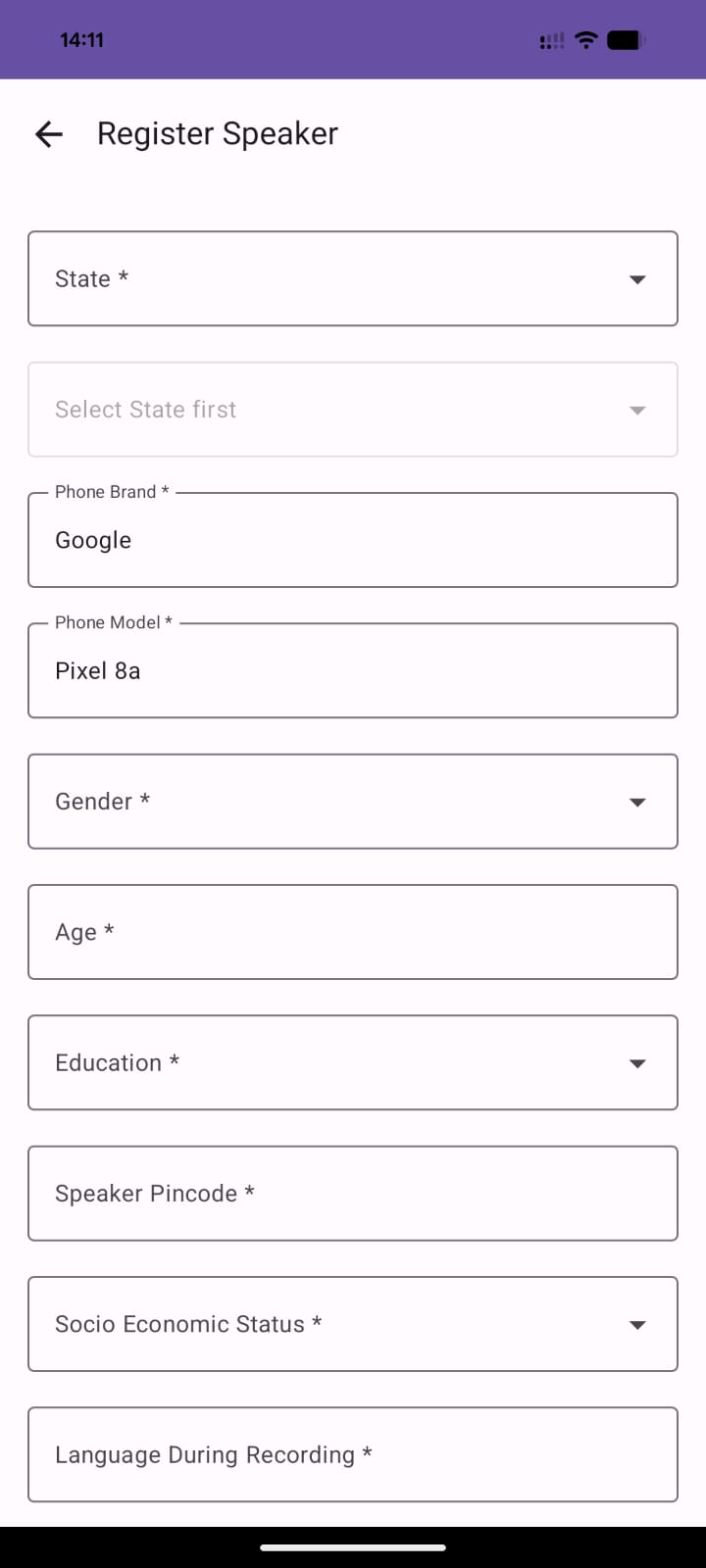}
        \caption{First}
    \end{subfigure}
    \hfill
    \begin{subfigure}{0.23\textwidth}
        \includegraphics[width=\textwidth]{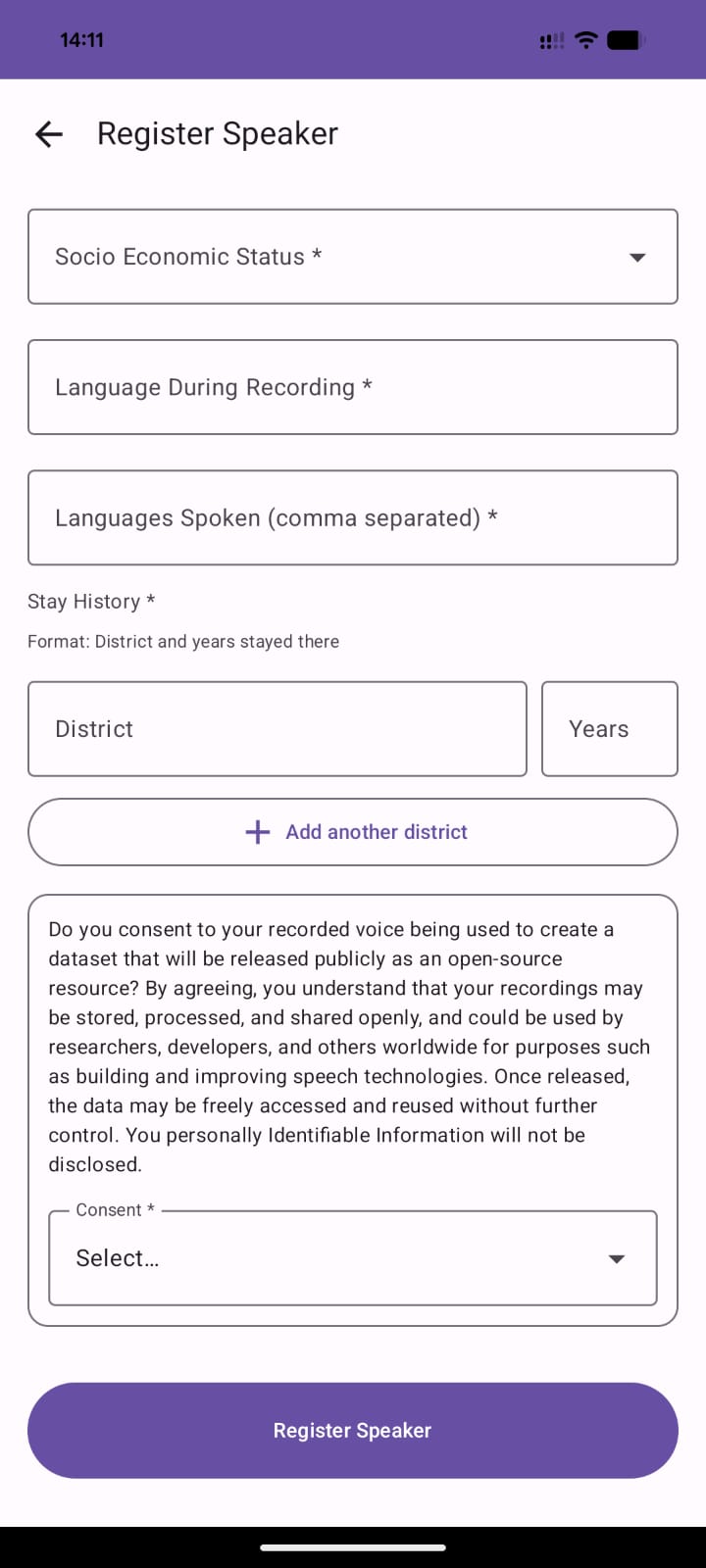}
        \caption{Second}
    \end{subfigure}
    \hfill
    \begin{subfigure}{0.23\textwidth}
        \includegraphics[width=\textwidth]{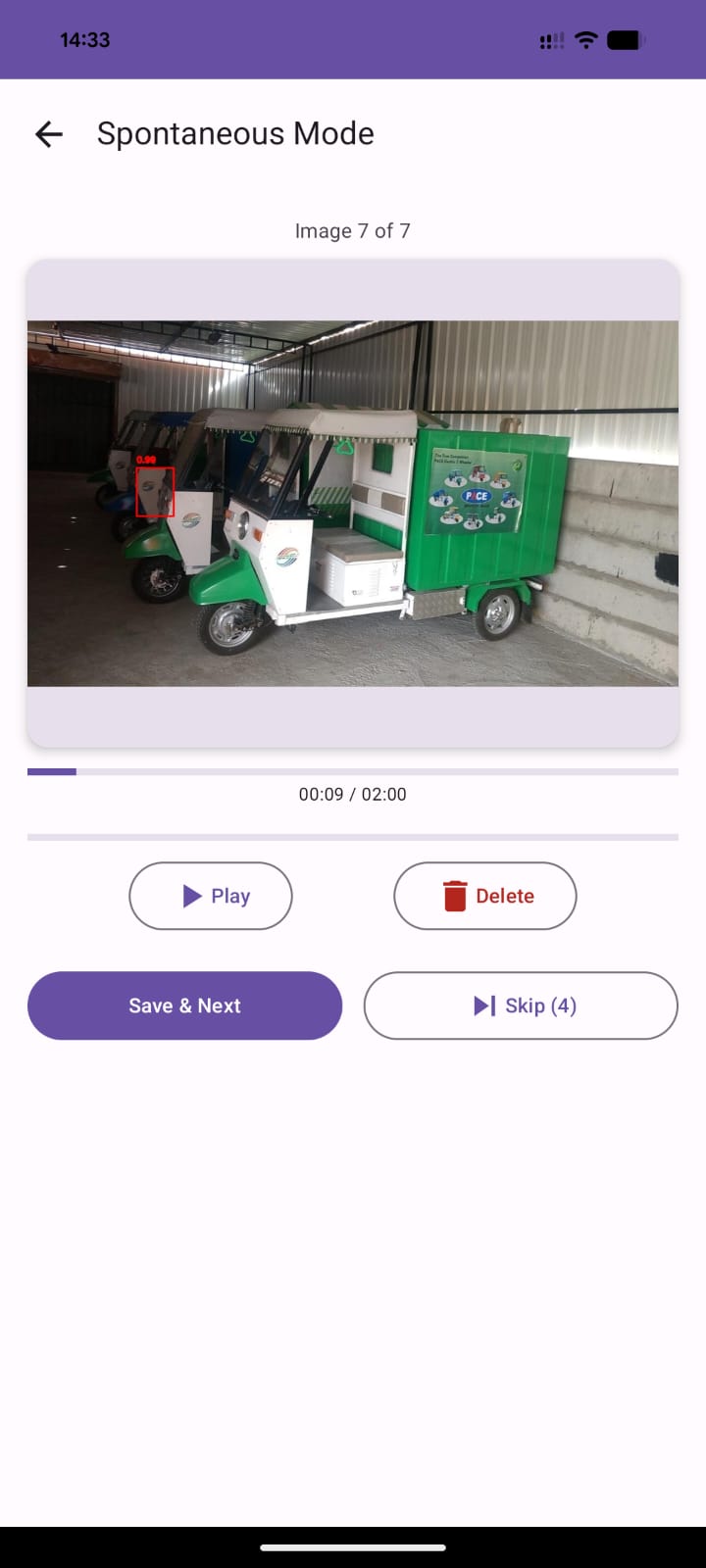}
        \caption{Third}
    \end{subfigure}
    \hfill
    \begin{subfigure}{0.23\textwidth}
        \includegraphics[width=\textwidth]{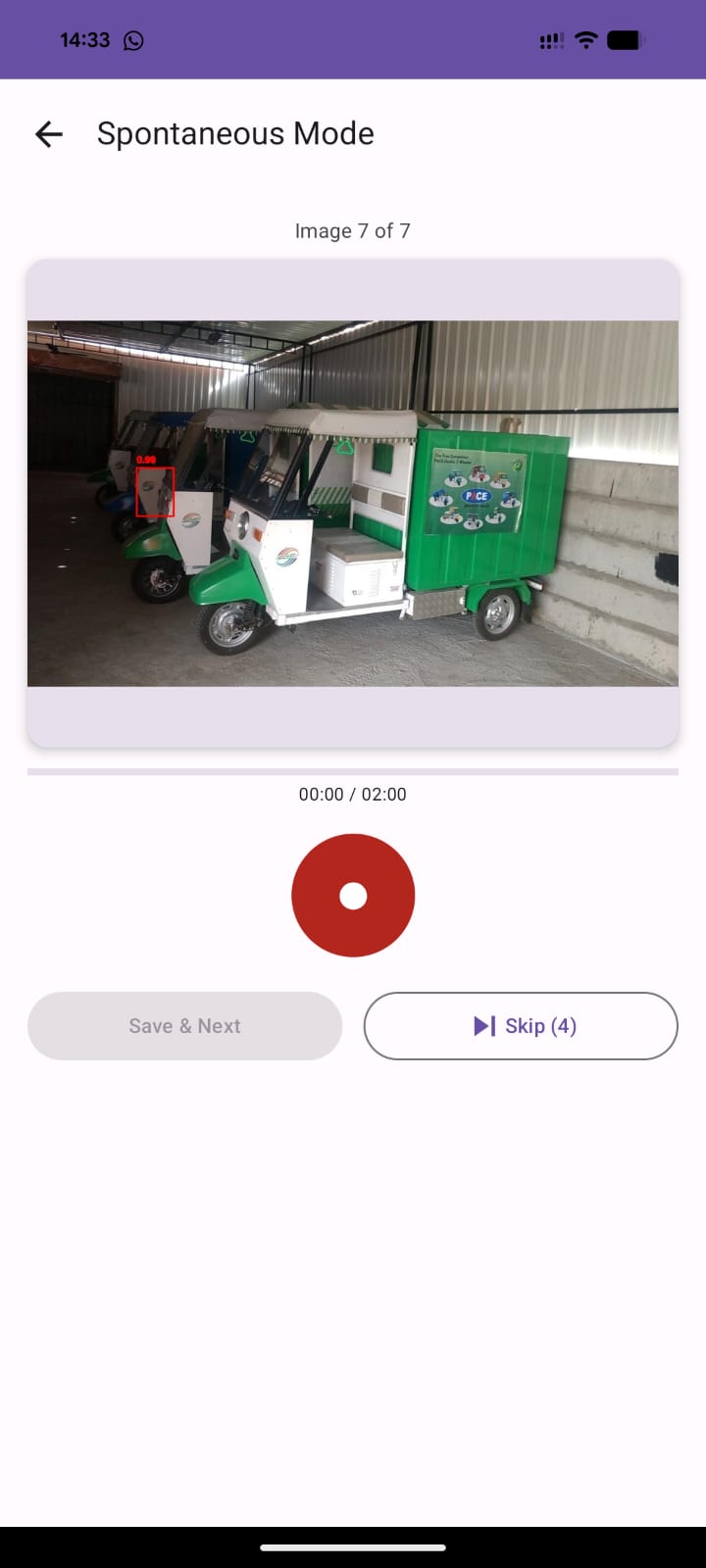}
        \caption{Fourth}
    \end{subfigure}
\caption{Reference images from the sample audio collection application used for data collection.}
\end{figure}

\subsection{Format of Deliverables}
\label{sec:deliverables}

\subsubsection{Folder Structure}
\label{sec:folder-structure}

\noindent Folder names are shown in angle brackets \texttt{<\,>}; all other entries are files. The deliverable is organized hierarchically by district, then by speaker:

\begin{itemize}
    \item \texttt{<District1>}
    \begin{itemize}
        \item \texttt{<Speaker1>}
        \begin{itemize}
            \item \textbf{Audio files:} \texttt{AudioFileForImage1}, \texttt{AudioFileForImage2}, \ldots\ (one audio recording per image prompt).
            \item \textbf{Per-audio TSV info files:} one file per audio recording (\texttt{AudioFileForImage1\_info.tsv}, \texttt{AudioFileForImage2\_info.tsv}, \ldots), each containing the following tab-separated fields: \texttt{IMAGE\_FILENAME}, \texttt{AudioFileName}, \texttt{AudioSegmentID}, \texttt{SegmentStartTime}, \texttt{SegmentEndTime}.
            \item \textbf{Speaker metadata file:} containing demographic and recording-context information for the speaker.
            \item \textbf{Transcription file:} a TSV file containing the following tab-separated fields: \texttt{TranscriberID}, \texttt{IMAGE\_FILENAME}, \texttt{AudioFileName}, \texttt{AudioSegmentID}, \texttt{SegmentStartTime}, \texttt{SegmentEndTime}, \texttt{Transcription}.
        \end{itemize}
        \item \texttt{<Speaker2>}, \texttt{<Speaker3>}, \ldots\ (same structure as \texttt{<Speaker1>}).
        \item \textbf{Transcriber metadata files:} one file per transcriber, e.g., \texttt{Transcriber1\_metadata.txt}, \texttt{Transcriber2\_metadata.txt}, \ldots
    \end{itemize}
    \item \texttt{<District2>}, \texttt{<District3>}, \ldots\ (same structure as \texttt{<District1>}).
\end{itemize}
\subsubsection{Unicode}
The Unicodes to be used are listed at \href{http://languagelog.ldc.upenn.edu/myl/ldc/IndianScriptsUnicode.html}{Indian Scripts Unicode reference}.

\subsubsection{Reference Files}
For each speaker,  we expect 5 reference files. These are files which are guaranteed by the vendor/company to be belonging to the speaker. These must be sourced from the ground at the time of recording.

\subsubsection{Single-Batch Delivery per Speaker}
All files for a particular speaker must be delivered at once only. Delivery of files for one speaker across batches is not permitted. If files for a speaker are received  across more than one batch, the speaker data beyond its first delivery batch will be rejected.

\subsubsection{Minimum Acceptable Files per Speaker}
Each speaker must have a minimum of 10 acceptable files as per the specifications outlined in this document. We shall discard files for those speakers where this condition is violated.

%==============================================================
\subsection{Part 3:Acceptance Criteria}
\addcontentsline{toc}{section}{Part 3: Acceptance Criteria}
\subsubsection{ Audio}
\begin{itemize}
    \item[3.1.1] The audio data must follow the specification given above.
    \item[3.1.2] If less than 3\% of the audio in a batch is found to violate the specifications (listed in Parts 1, 2, and 3 above, as indicatively detected through programmatic checks of the university in as-is condition), then the entire batch will be accepted and paid for in entirety.
    \item[3.1.3] If more than 10\% of audio in a batch delivered is found to violate the specifications (listed in Parts 1, 2, and 3 above), the entire batch will be rejected.
    \item[3.1.4] If less than 10\% but more than 3\% of audio in a batch delivered is found to violate the specifications, audio files of equivalent duration (location and speaker profile matched to maintain desired age, gender, and location distribution) should be re-recorded and delivered.
    \item[3.1.5] Each speaker must have audio files corresponding to a minimum of 40 images post the entire audio approval process. A speaker failing this criterion shall be rejected entirely.
\end{itemize}
\noindent\textit{The quantity will always be measured after removing silence portions within the files.}

\subsubsection{Transcription}
\begin{itemize}
    \item[3.2.1] The transcription must follow the guidelines as provided by the University.
    \item[3.2.2] If less than 10\% of transcription in a batch delivered is found to violate the specifications, they should be re-transcribed.
    \item[3.2.3] If more than 10\% of transcription in a batch delivered is found to violate the specifications, the entire batch will be rejected.
\end{itemize}

\begin{figure}[htbp]                  % use figure* for two-column papers
    \centering
    \resizebox{0.85\textwidth}{!}{%
    \begin{tikzpicture}[
        font=\sffamily\footnotesize,
        node distance=0.8cm,
        process/.style={
            rectangle, draw, thin,
            minimum width=2.4cm, minimum height=1.1cm,
            align=center, text width=2.2cm, inner sep=2pt
        },
        terminator/.style={
            ellipse, draw, thin,
            minimum width=2.6cm, minimum height=0.9cm,
            align=center, inner sep=2pt
        },
        data/.style={
            trapezium, trapezium left angle=70, trapezium right angle=110,
            draw, thin,
            minimum width=2.6cm, minimum height=0.9cm,
            align=center, text width=2.4cm, inner sep=2pt,
            trapezium stretches body
        },
        arrow/.style={-{Stealth[length=2mm]}, thin}
    ]
    % ---- LEFT MAIN COLUMN ----
    \node[terminator] (vendor) {Vendor Delivers data};
    \node[process, below=of vendor] (initial) {Initial\\Automated\\Checks by\\University};
    \node[process, below=of initial] (speaker) {Speaker List\\generation by\\University};
    \node[process, below=of speaker] (refprov) {Reference files\\provided by\\vendor};
    \node[process, below=of refprov] (refcheck) {Reference\\Files Check by\\University};
    \node[process, below=of refcheck] (audioqc) {Automated\\Audio QC};
    \node[process, below=of audioqc] (finalaudio) {Final Audio\\QC\\(Post Manual\\QC)};
 
    \node[data, right=1.6cm of initial] (feedback) {Feedback on Initial Checks to vendor};
    \node[data, left=1.4cm of audioqc] (indicaudio) {Indicative Audio Results};
    \node[data, left=1.4cm of finalaudio] (finalauto) {Final Auto Results};
    \node[data, right=1.4cm of audioqc] (chunks) {Transcription Chunks sent to vendors};
 
    % ---- RIGHT BRANCH (Transcription) ----
    \node[terminator, below=of chunks] (transdeliv) {Transcription Data\\Delivery by Vendor};
    \node[process, below=of transdeliv] (transinit) {Transcription\\Initial\\Automated\\QC};
    \node[process, below=of transinit] (transfinal) {Final Transcription\\QC (Post Manual\\QC)};
 
    \node[data, right=1.4cm of transinit] (indictrans) {Indicative Transcription Results};
    \node[data, right=1.4cm of transfinal] (finaltrans) {Final Transcription Result};
 
    % ---- ARROWS ----
    \draw[arrow] (vendor) -- (initial);
    \draw[arrow] (initial) -- (speaker);
    \draw[arrow] (speaker) -- (refprov);
    \draw[arrow] (refprov) -- (refcheck);
    \draw[arrow] (refcheck) -- (audioqc);
    \draw[arrow] (audioqc) -- (finalaudio);
 
    \draw[arrow] (initial) -- (feedback);
    \draw[arrow] (audioqc) -- (indicaudio);
    \draw[arrow] (finalaudio) -- (finalauto);
    \draw[arrow] (audioqc) -- (chunks);
 
    \draw[arrow] (chunks) -- (transdeliv);
    \draw[arrow] (transdeliv) -- (transinit);
    \draw[arrow] (transinit) -- (transfinal);
    \draw[arrow] (transinit) -- (indictrans);
    \draw[arrow] (transfinal) -- (finaltrans);
    \end{tikzpicture}
    }%
    \caption{Operational workflow for audio and transcription data exchange between vendors and the university.}
    \label{fig:operational_pipeline}
\end{figure}

% =============================================================
% Required packages (standard, usually already loaded):
%   \usepackage{longtable}
%   \usepackage{array}
%   \usepackage{booktabs}
% =============================================================

\section{Audio QC Guidelines}
\label{sec:guidelines-overview}

\noindent Data quality executive,  will be required to listen to the audio in videos and answer specific questions based on what they hear. The questions  are designed to help them to assess the quality and relevance of the audio. 

%==============================================================
\subsection*{Questions}

\renewcommand{\arraystretch}{1.3}

\begin{longtable}{@{}
    >{\raggedright\arraybackslash}p{0.32\linewidth}
    >{\raggedright\arraybackslash}p{0.22\linewidth}
    >{\raggedright\arraybackslash}p{0.40\linewidth}
@{}}
\toprule
\textbf{Field} & \textbf{What to Check} & \textbf{Expected Response} \\
\midrule
\endfirsthead

\toprule
\textbf{Field} & \textbf{What to Check} & \textbf{Expected Response} \\
\midrule
\endhead

\bottomrule
\endfoot

Do you hear any humans speaking in this video?
& Are humans speaking in the audio?
& \textbf{Yes}: If a human is speaking.\newline
  \textbf{No}: If there is only background noise or blank audio. \\

Do you hear only one person speaking?
& Is only one person speaking?
& \textbf{Yes}: If only one person is speaking.\newline
  \textbf{No}: If there is more than one person, or if other sounds (e.g., people talking, baby crying) impact the clarity of the speaker's voice. \\

Who do you think is speaking in the audio?
& Identify the speaker's gender or age.
& Options: \textbf{Male}, \textbf{Female}, \textbf{Junior}, \textbf{NA}.\newline
  \textit{Impersonation}: If the speaker is impersonating a female or a child, select accordingly. \\

Are you able to understand what you hear?
& Can you understand the audio?
& \textbf{Yes}: If the entire sentence is understandable.\newline
  \textbf{No}: If you can only partially understand or cannot understand the whole sentence. \\

Does it sound like somebody from your district?
& Is the accent or dialect familiar?
& \textbf{Yes}: If the dialect and accent match those spoken in your district.\newline
  \textbf{No}: If the speaker is using a different dialect or language. \\

Does the audio contain disturbance that makes it difficult to understand what is spoken?
& Is there noise or disturbance in the audio?
& \textbf{No}: If there is no disturbance or only slight noise, but the speaker's voice is still understandable.\newline
  \textbf{Yes}: If the speaker's voice is unclear due to loud background noise. \\

Does the audio sound like a complete sentence?
& Is the audio a complete sentence?
& \textbf{Yes}: If it is a complete sentence.\newline
  \textbf{No}: If it is an incomplete sentence. \\

Does this audio talk about the image?
& Is the audio relevant to the image?
& \textbf{Yes}: If the audio describes the image.\newline
  \textbf{No}: If the audio is not related to the image. \\

Do you find any personal information about either the person in the audio or the image?
& Does the content contain sensitive information?
& \textbf{yes-audio}: If the audio contains sensitive information (e.g., PAN card, Aadhaar card, ATM card number, bank details).\newline
  \textbf{yes-image}: If the image contains personal information and the face is not blurred.\newline
  \textbf{yes-both}: If both the audio and image contain personal information.\newline
  \textbf{No}: If no sensitive information is present and the face is blurred. \\

Identify whether the spoken language is the same as the mentioned one?
& Does the language match the mentioned one?
& \textbf{Yes}: If the spoken language matches the one mentioned in the `language' column.\newline
  \textbf{No}: If it is a different language. \\

Is there any distortion in the image you see in the audio?
& Is the image clear or distorted?
& \textbf{Yes}: If the image is blurry, unclear, or altered.\newline
  \textbf{No}: If the image is clear. \\
\end{longtable}

%==============================================================
\subsection{Frequently Asked Questions (FAQs) prepared for QC process}

\noindent Below are common questions that can help while conducting audio quality checks, along with detailed explanations to help the QC team to understand how to approach each task. 
%--------------------------------------------------------------
\paragraph{1. Do you hear any humans speaking in this video?}
\textbf{What to check:}
\begin{itemize}
    \item You should hear a human speaking clearly in the audio.
    \item If you can hear and understand the person speaking despite the background noise, answer \textbf{Yes}. The key is whether the human voice is clear and understandable.
    \item If there is no human voice and you hear only other noises (vehicles, animals, background noise), answer \textbf{No} and add the comment \textit{``No human is speaking, only background noise''}.
    \item If the audio is blank, answer \textbf{No} and add the comment \textit{``Blank Audio''}.
\end{itemize}

%--------------------------------------------------------------
\paragraph{2. Do you hear only one person speaking?}
\textbf{What to check:}
\begin{itemize}
    \item Pay attention to the number of distinct voices in the audio. If you hear just one person speaking clearly, answer \textbf{Yes}.
    \item If one person is speaking clearly and another voice is just in the background, still answer \textbf{Yes}.
    \item If two people are speaking, answer \textbf{No}.
    \item If background noise (two or more people talking, kids talking, baby crying, someone speaking on the mic) is impacting the speaker's voice (the one describing the image), answer \textbf{No}. Comment: \textit{``More than one person speaking''}.
\end{itemize}

%--------------------------------------------------------------
\paragraph{3. Who do you think is speaking in the audio?}
\textbf{What to check:}
\begin{itemize}
    \item Listen carefully to the pitch, tone, and overall sound of the voice. Use your best judgment to determine if the voice belongs to a male, female, or child.
    \item If you are unsure between female and child, lean towards selecting \textbf{Female}.
    \item If unsure between male and female, go with your initial impression.
    \item If someone is impersonating a feminine or a kid's voice, answer as per the voice. Comment: \textit{``impersonating feminine voice''}.
\end{itemize}

%--------------------------------------------------------------
\paragraph{4. Are you able to understand what you hear?}
\textbf{What to check:}
\begin{itemize}
    \item Answer \textbf{Yes} if you understand the whole sentence.
    \item Answer \textbf{No} if you do not understand or only partially understand.
    \item If the audio is muffled or unclear to the point where you cannot understand the words or the meaning, answer \textbf{No}.
\end{itemize}

%--------------------------------------------------------------
\paragraph{5. Does it sound like somebody from your district?}
\textbf{What to check:}
\begin{itemize}
    \item Listen for the local dialect, accent, and language that matches your district. If the language and accent sound familiar and local to your area, answer \textbf{Yes}.
    \item If you are unsure, inform your POC.
    \item If the language or accent does not sound like it is from your district, answer \textbf{No} and add the comment \textit{``not a local language in my district''}.
\end{itemize}

%--------------------------------------------------------------
\paragraph{6. Is the audio clear and without any disturbances?}
\textbf{What to check:}
\begin{itemize}
    \item Answer \textbf{Yes} if the speaker's voice is clear and dominant and can be understood over any background noise.
    \item Answer \textbf{No} if the background noise is so loud that it drowns out the speaker's voice or makes it hard to understand. Comment: \textit{``very loud background noise''}.
\end{itemize}

%--------------------------------------------------------------
\paragraph{7. Does the audio sound like a complete sentence?}
\textbf{What to check:}
\begin{itemize}
    \item Answer \textbf{Yes} if the sentence is complete.
    \item If the sentence is complete and understandable, even if it is short, answer \textbf{Yes}.
    \item Answer \textbf{No} if the sentence is cut off or does not make complete sense. Comment: \textit{``Incomplete Sentence''}.
\end{itemize}

\noindent \textbf{Note:} If an audio ends with a near-final consonant where the trailing vowel is faint but recognizable (e.g., the sound of \emph{h} present but the full \emph{hai}/\emph{hain} not fully voiced), it will still be considered a complete sentence.

%--------------------------------------------------------------
\paragraph{8. Does this audio talk about the image?}
\textbf{What to check:}
\begin{itemize}
    \item Answer \textbf{Yes} if the audio describes or is related to the image shown.
    \item Answer \textbf{No} if the audio does not describe the image. Comment: \textit{``image not related to audio''}.
\end{itemize}

%--------------------------------------------------------------
\paragraph{9. Do you find any personal information about either the person in the audio or the image?}
\textbf{What to check:}
\begin{itemize}
    \item Answer \textbf{No} if there is no personal information.
    \item If a person is posing for the camera and the face is blurred, answer \textbf{No}.
    \item Answer \textbf{yes-audio} if there is personal information in the audio (sensitive information like bank details, Aadhaar card number, OTP, PAN card details, etc.).
    \item If the image has personal information (face is not blurred) \textit{and} sensitive information is being shared in the audio (OTP, bank details, etc.), answer \textbf{yes-both}.
\end{itemize}

\noindent \textbf{Cases (yes-image):}
\begin{itemize}
    \item A person poses for the camera and the face is not blurred.
    \item Two or more people pose for the camera and no one's face is blurred.
    \item People pose for the camera in public places and faces are visible.
    \item Two people take a selfie where one face is blurred and the other is not.
\end{itemize}

\noindent \textbf{Comment:} \textit{Audio or image contains personal information of someone.}

%--------------------------------------------------------------
\paragraph{10. Identify whether the spoken language is the same as the mentioned one?}
\textbf{What to check:}
\begin{itemize}
    \item If the audio is in the same language mentioned in the language column, answer \textbf{Yes}.
    \item If the audio is in a different language than the one mentioned, answer \textbf{No}. Comment: \textit{``Language mismatch''}.
\end{itemize}
\section{Transcription Guidelines }
\label{sec:transcription_guidelines}

The following transcription guidelines are adapted from the Digital India Bhashini Project, which in turn are inspired by similar guidelines created by a commercial transcription agency and by NIST.

\subsection{General Principles}
\begin{itemize}
    \item Transcribe a word only if you can hear and understand it properly. If the spoken word/text cannot be understood due to the speaker’s manner of speech then mark it as [unintelligible]. On the other hand, if the spoken text cannot be heard due to poor recording, volume or noise then mark it as [inaudible]. An audio segment will be considered as invalid if more than 25\% of the words in that audio segment transcription (i.e., a sentence) have the [unintelligible] and/or [inaudible] mark.

\item Do not paraphrase the speech.

\item Do not correct grammatical errors made by the speakers.

\item In the case of misspoken words, transcribe the one with a spelling mistake followed by the correct spelling within curly braces. Example: If a speaker pronounces "remuneration" as “renumeration” then it should still be transcribed as “renumeration \{remuneration\}”.

\item Do not expand spoken short forms (e.g., ain’t, don’t, can’t, it should be retained as it is)
\item Retain colloquial slang as it is (e.g., gotcha, gonna, wanna, etc).

\item If the spoken words belong to a language that the transcriber does not know then it should be tagged as <UNKNOWN\_SEGMENT> with the appropriate timestamps. If transcription of a sentence contains <UNKNOWN\_SEGMENT> for more than 25\% of the words, then it will not be accepted
\item Every pause more than 0.5 sec should be transcribed as <PAUSE>
\end{itemize}

\subsection{Names, titles, acronyms, punctuations, and numbers}
\begin{itemize}
\item Proper names should be transcribed in a case-sensitive manner in applicable languages (e.g., English). Initials should be in capital letters with no period following. For example: “M K Stalin would be sworn in as the Chief Minister”.
\item Titles and abbreviations are transcribed as words. For example: Dr. → Doctor except if the abbreviated form is pronounced as it is.  For example, if the speaker says “Apple Inc” (instead of “Apple Incorporate”), the word 'Inc' should be transcribed.
\item Punctuation marks should be used in transcription as appropriate. For example: "don't" 
Apart from the letters (or Aksharas) only the following special characters can be used - COMMA (,), FULLSTOP (.), QUESTION MARK (?), APOSTROPHE (‘). UNDERSCORE ( '\_' ) to be used only for acronym (see the next bullet). Purna viram (|) may be used to indicate end of a sentence only.

\item Numbers should be transcribed as they are spoken and normalized to word form (no numerals). For example: 16 → sixteen, 112 → one hundred and twelve.
Times of the day and dates: always capitalize AM and PM (wherever applicable). When using o'clock, spell out the numbers: eleven o'clock.
\end{itemize}
\subsection{Incomplete utterances}

These are utterances which are incomplete because the speaker forgot what he wanted to say or was stopped mid-way and corrected an error or was interrupted by someone. Indicate such utterances by putting a ‘--’ at the end of the utterance as opposed to a full-stop or question mark.

\subsection{Verbatim transcription}
The speech should be transcribed verbatim. 
However, the following rules should be used for transcribing errors made by the speakers.
Errors that should be transcribed as it is:
Speech errors: “I was in my office, no sorry, home” should be transcribed as it is.
Slang words: kinda, gonna, wanna, etc should be transcribed as it is.
Repetitions: “I have I have got the book” should be transcribed as it is.

Errors that should be transcribed:
False starts: “I, um, er, I was going to the mall” should be transcribed as it is.
Filler sounds: um, uh, er, hmm, etc. should be transcribed as it is.
Stutters: “I w-w-went t-t-to the mall” should be transcribed as it is. Following the guidelines for the SWITCHBOARD corpus there should be a hyphen between the stutters as shown in the above sentence.

\subsection{Non-speech (acoustic) events}

Foreground sounds made by the speaker should be transcribed. These include lip smacks, tongue clicks, inhalation and exhalation between words, yawning coughing, throat clearing, sneezing, laughing, chuckling, etc. The categories specified in Table 1 in the guidelines for the SWITCHBOARD corpus \href{(https://isip.piconepress.com/projects/switchboard/doc/transcription_guidelines/transcription_guidelines.pdf)}, should be used.

% Requires in preamble:
%   \usepackage{graphicx}
%   \usepackage{subcaption}

\section{Speaker Demographics}
\label{sec:demographics}

\begin{figure*}[t]
    \centering
    % --------- Row 1 ---------
    \begin{subfigure}[t]{0.48\linewidth}
        \centering
        \includegraphics[width=\linewidth]{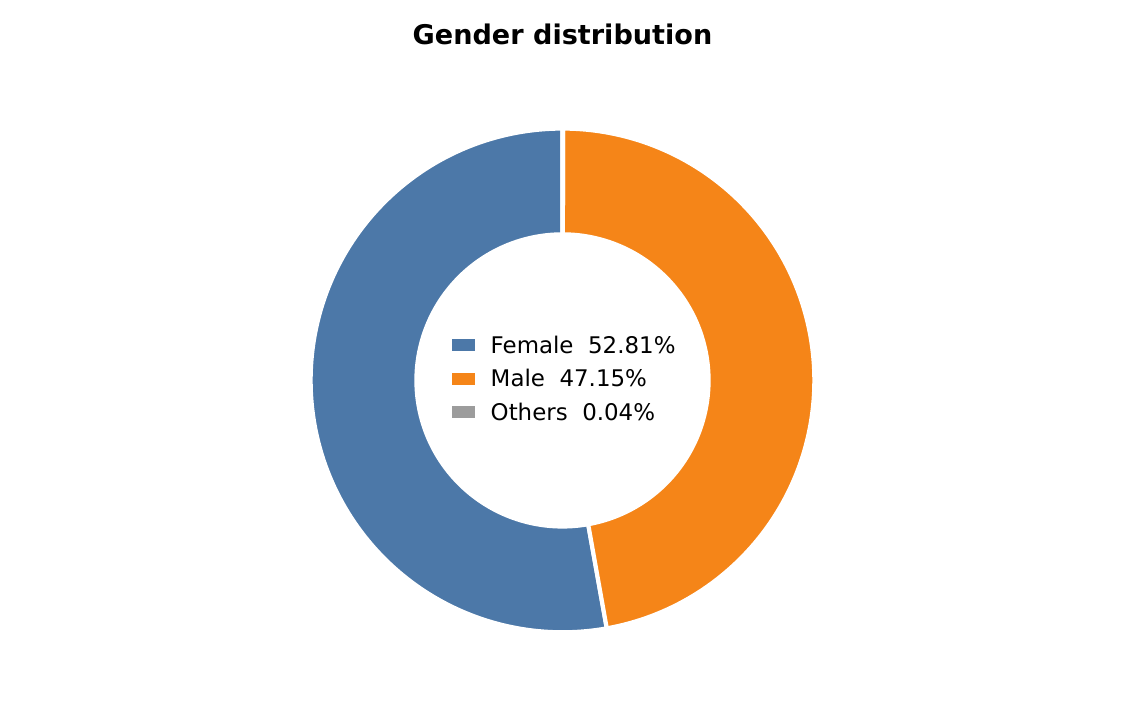}
        \caption{Gender.}
        \label{fig:demo-gender}
    \end{subfigure}
    \hfill
    \begin{subfigure}[t]{0.48\linewidth}
        \centering
        \includegraphics[width=\linewidth]{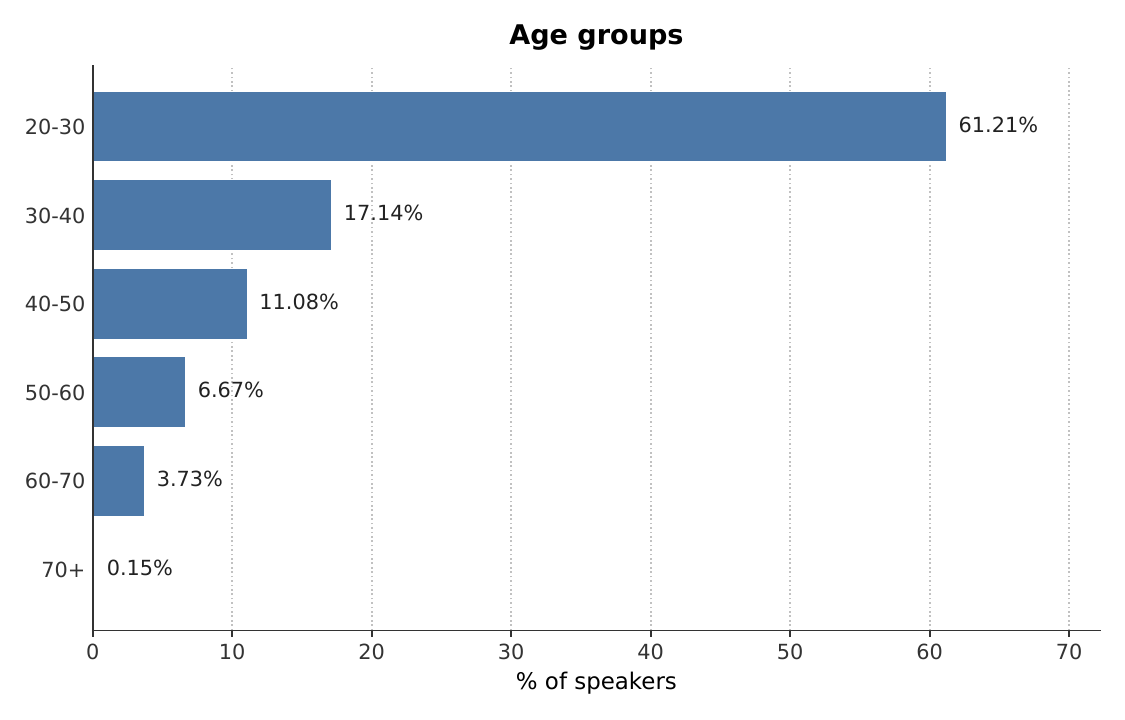}
        \caption{Age group.}
        \label{fig:demo-age}
    \end{subfigure}

    \vspace{0.8em}

    % --------- Row 2 ---------
    \begin{subfigure}[t]{0.48\linewidth}
        \centering
        \includegraphics[width=\linewidth]{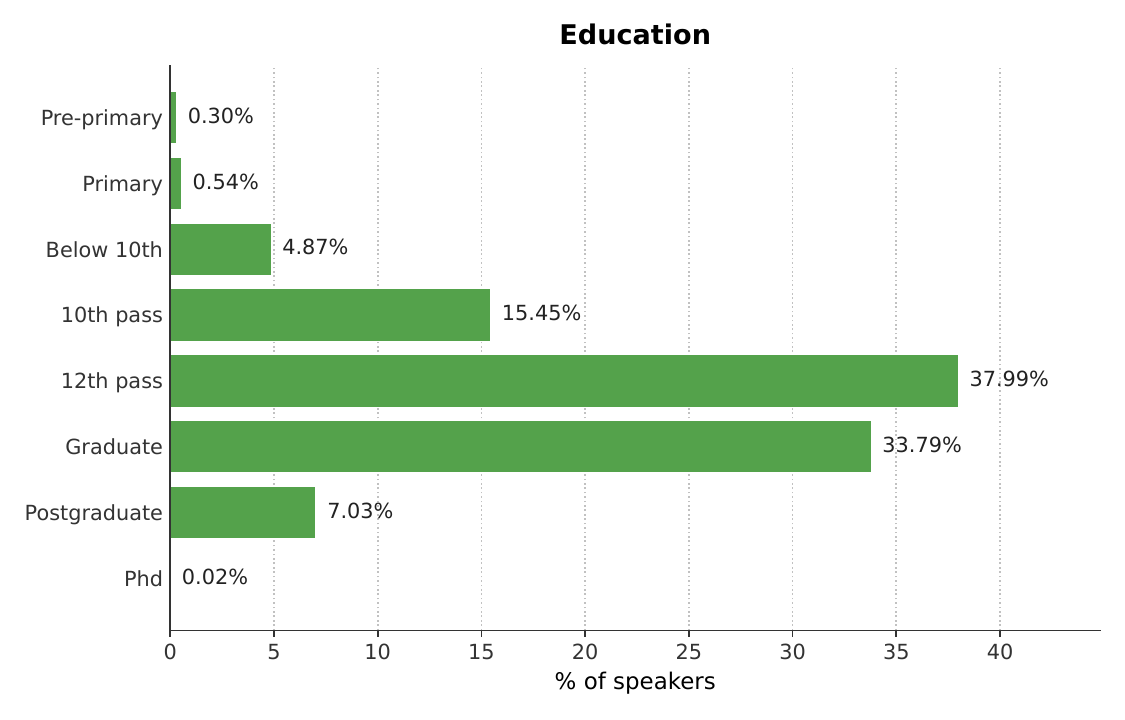}
        \caption{Education.}
        \label{fig:demo-education}
    \end{subfigure}
    \hfill
    \begin{subfigure}[t]{0.48\linewidth}
        \centering
        \includegraphics[width=\linewidth]{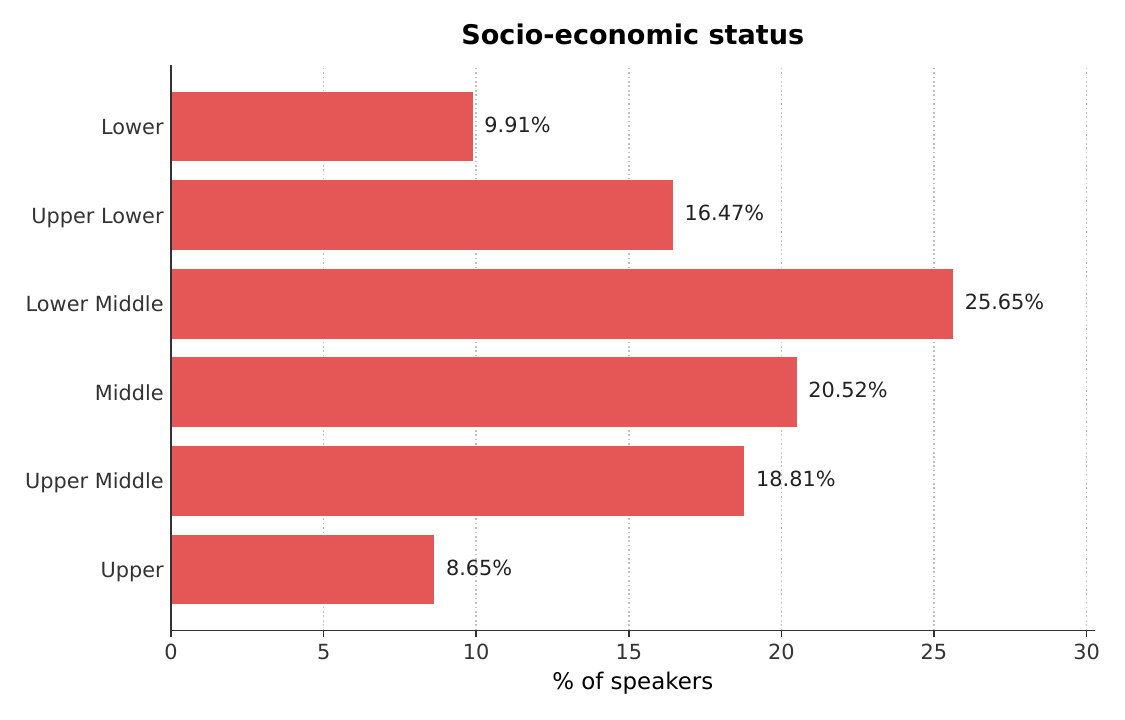}
        \caption{Socio-economic status.}
        \label{fig:demo-socio}
    \end{subfigure}

    \caption{Demographic composition of the enrolled speaker pool across gender, age, education, and self-reported socio-economic status.}
    \label{fig:demographics}
\end{figure*}

This section reports the demographic composition of the 158,441 speakers and Figure~\ref{fig:demographics} summarises the four primary demographic dimensions: gender, age, education, and self-reported socio-economic status.

\paragraph{Gender.}
The enrolled pool is near-balanced, with a slight female majority: 52.81\% female, 47.15\% male, and 0.04\% identifying as other (Figure~\ref{fig:demo-gender}).

\paragraph{Age.}
Speakers skew strongly young (Figure~\ref{fig:demo-age}). The 20--30 bracket alone accounts for 61.21\% of the pool, with a further 17.14\% in the 30--40 bracket. Representation thins progressively in older groups: 40--50 (11.08\%), 50--60 (6.67\%), 60--70 (3.73\%), and 70+ (0.15\%).

\paragraph{Education.}
The pool is dominated by speakers with secondary and undergraduate education (Figure~\ref{fig:demo-education}): 37.99\% are 12th-pass and 33.79\% are graduates, with 10th-pass speakers contributing a further 15.45\%. The remaining ${\sim}13$\% is split between postgraduate, below-10th, and primary-level speakers.

\paragraph{Socio-economic status.}
Self-reported socio-economic status is concentrated in the middle strata (Figure~\ref{fig:demo-socio}). Lower-Middle (25.65\%) is the modal class, followed by Middle (20.52\%) and Upper-Middle (18.81\%); together, these three strata account for almost two-thirds of the pool. The tails are covered by Upper-Lower (16.47\%), Lower (9.91\%), and Upper (8.65\%).

\medskip

\noindent Overall, the enrolled pool is gender-balanced and skews young and middle-strata in education and economic background, with thinner but non-negligible representation in the older, less-educated, and economically peripheral tails.
% Required in preamble:
%   \usepackage{longtable}
%   \usepackage{booktabs}
%   \usepackage{array}
%
% NOTE: This file uses custom column types `N`, `C`, and `D`. These must
% be defined in your preamble. A typical definition is:
%   \newcolumntype{N}{>{\raggedright\arraybackslash}p{2.5cm}}
%   \newcolumntype{C}{>{\centering\arraybackslash}p{0.7cm}}
%   \newcolumntype{D}{>{\raggedright\arraybackslash}p{1.4cm}}
%
% NOTE: The two grand totals reported below differ slightly
% (31255.10 vs 31245.34 hrs). Preserved verbatim from the source —
% please reconcile before final submission if intended to match.

\section{District-wise and Language-wise Data Distribution}
\label{sec:district-language-distribution}
Table \ref{tab:districts} shows the district-wise distribution of the released data, including the number of languages recorded in each district along with the audio and transcription duration. Table \ref{tab:languages} presents the language-wise distribution, showing the number of districts in which each language was collected, along with the corresponding audio and transcription durations.
%==============================================================
% Table 1: District-wise distribution
%==============================================================

\begin{longtable}{N C D D | N C D D}
\caption{Audio collection and transcription duration per district (hours), with the number of distinct languages recorded in each district.}
\label{tab:districts} \\
\toprule
\textbf{District} & \textbf{Langs} & \textbf{Audio} & \textbf{Transc.} & \textbf{District} & \textbf{Langs} & \textbf{Audio} & \textbf{Transc.} \\
                  &                & \textbf{(hrs)} & \textbf{(hrs)}   &                  &                & \textbf{(hrs)} & \textbf{(hrs)}   \\
\midrule
\endfirsthead

\multicolumn{8}{c}{\tablename\ \thetable\ -- \textit{continued from previous page}} \\
\toprule
\textbf{District} & \textbf{Langs} & \textbf{Audio} & \textbf{Transc.} & \textbf{District} & \textbf{Langs} & \textbf{Audio} & \textbf{Transc.} \\
                  &                & \textbf{(hrs)} & \textbf{(hrs)}   &                  &                & \textbf{(hrs)} & \textbf{(hrs)}   \\
\midrule
\endhead

\midrule \multicolumn{8}{r}{\textit{continued on next page}} \\
\endfoot

\midrule
\multicolumn{8}{r}{\textbf{Grand Total --- Audio: 31255.45 hrs \quad Transcription: 2043.31 hrs}} \\
\bottomrule
\endlastfoot

Adilabad           &  2 &  12.42 &  0.98 & Kollam             &  2 & 207.74 & 23.71 \\
Aizawl             & 10 & 220.00 & 23.94 & KomaramBheem       &  4 &  28.01 &  2.46 \\
Alipurduar         &  5 & 212.23 & 20.43 & Koppal             &  3 & 207.93 & 20.59 \\
Anantpur           &  5 & 218.30 & 11.14 & Korba              &  2 & 218.63 &  9.32 \\
Annamaya           &  3 &  72.12 &  9.18 & Koriya             &  4 &  15.17 &  0.00 \\
Araria             &  5 & 213.74 & 10.19 & Kozhikode          &  2 &  47.79 &  6.07 \\
Aurangabad         &  3 & 191.27 &  9.85 & Krishna            &  6 & 185.92 &  9.67 \\
Balrampur          &  3 & 208.21 & 12.81 & Lakhisarai         &  6 & 209.62 &  9.49 \\
Bangalore          & 13 & 208.69 & 19.60 & Lalitpur           &  2 & 220.00 & 16.89 \\
Barmer             &  3 &  54.00 &  0.62 & Longding           &  3 & 210.47 & 20.36 \\
Bastar             &  7 & 208.83 & 10.11 & LowerDibangvalley  &  6 & 227.76 & 17.44 \\
Begusarai          &  6 & 221.46 &  8.18 & Lucknow            &  3 & 278.48 & 22.21 \\
Belgaum            &  6 & 201.21 & 12.68 & Madhepura          &  6 & 212.05 &  7.91 \\
Bellary            &  6 & 201.13 & 11.87 & Mahabubabad        &  3 & 215.61 & 21.29 \\
Bhagalpur          &  8 & 220.13 &  8.57 & Malda              &  3 & 203.96 & 11.02 \\
Bhopal             &  4 & 220.00 & 18.58 & Malkangiri         & 11 & 208.76 & 21.57 \\
Bidar              &  3 & 201.96 & 20.01 & Manyam             &  3 &  47.09 &  5.75 \\
Bijapur            &  4 & 189.85 &  8.31 & Mumbaisuburban     &  4 & 202.30 & 10.24 \\
Bikaner            &  3 & 211.12 & 19.71 & Muzaffarnagar      &  6 & 216.13 & 11.13 \\
Bilaspur           &  5 & 220.00 & 11.15 & Muzaffarpur        &  6 & 219.24 & 10.84 \\
Budaun             &  8 & 220.00 & 11.87 & Mysore             &  6 & 205.04 & 12.46 \\
Chamarajanagar     &  5 & 203.98 &  9.04 & Nagaur             &  7 & 207.10 & 15.23 \\
Chandigarh         &  4 & 206.74 & 21.33 & Nagpur             &  4 & 196.12 & 10.99 \\
Chandrapur         &  3 & 195.57 &  9.00 & Nalgonda           &  5 & 215.66 &  8.47 \\
CharkhiDadri       &  3 & 195.72 & 13.87 & Namakkal           &  1 & 220.00 &  3.00 \\
Chennai            &  1 & 204.03 &  1.70 & Narayanpur         &  4 &  44.00 &  0.13 \\
Chittoor           &  6 & 222.20 & 10.73 & Navsari            &  3 &  23.56 &  0.42 \\
Churu              & 11 & 208.62 & 14.00 & NewDelhi           &  5 & 205.15 & 10.42 \\
CoochBehar         &  3 & 202.57 & 19.15 & Nilgiris           &  2 & 206.57 & 11.38 \\
DakshinaKannada    &  7 & 197.23 & 10.07 & North24Parganas    &  3 & 220.00 & 12.62 \\
DakshinDinajpur    &  2 & 220.00 & 10.73 & NorthGaroHills     &  3 &  50.94 &  4.63 \\
Darbhanga          &  3 & 208.21 &  8.80 & NorthSouthGoa      &  6 & 187.76 & 14.00 \\
Darjeeling         &  4 & 203.50 & 21.06 & Nuapada            &  5 & 120.70 & 12.56 \\
Deoghar            &  3 & 203.36 & 19.51 & Palamu             &  5 & 217.63 & 20.73 \\
Deoria             &  6 & 207.17 &  9.68 & PapumPare          &  6 & 210.84 & 14.68 \\
DevbhoomiDwarka    &  2 &  35.17 &  0.00 & PaschimMedinipur   &  2 & 220.00 &  8.47 \\
Dhalai             &  4 & 246.89 & 24.22 & Pathankot          &  5 & 218.44 & 23.38 \\
Dhar               &  4 & 220.00 & 18.97 & Patna              &  7 & 203.14 & 22.11 \\
Dharwad            &  6 & 208.97 & 11.07 & Pune               &  3 & 211.53 & 13.58 \\
Dhule              &  4 & 198.40 & 11.58 & Purnia             &  5 & 220.00 & 10.50 \\
Dimapur            & 15 & 210.94 & 16.46 & Purulia            &  4 & 212.69 &  9.65 \\
EastChamparan      &  6 & 212.10 &  8.58 & Raichur            &  5 & 209.87 & 11.16 \\
Etah               &  4 & 218.28 & 11.52 & Raigarh            &  3 & 205.03 & 10.80 \\
Fazilka            &  4 & 207.18 & 21.77 & Raipur             &  5 & 202.12 & 15.54 \\
Gandhinagar        &  5 & 128.46 &  9.59 & Rajnandgaon        &  4 & 208.60 &  9.83 \\
Gangtok            &  7 & 216.25 & 18.12 & Ranchi             &  4 & 251.27 & 21.11 \\
Garhwa             &  4 & 220.00 & 20.87 & Rohtak             &  3 & 200.50 & 16.64 \\
Gaya               &  6 & 226.20 &  8.98 & Saharanpur         &  2 & 200.19 & 17.79 \\
Ghazipur           &  8 & 208.13 &  8.26 & Saharsa            &  5 & 206.65 &  9.05 \\
Gondia             &  4 & 194.08 & 11.09 & Sahebganj          & 11 & 209.16 & 10.57 \\
Gopalganj          &  4 & 220.00 & 10.94 & Samastipur         &  7 & 220.73 & 11.42 \\
Gorakhpur          &  4 & 213.73 &  9.41 & Sambalpur          &  7 & 196.45 & 11.54 \\
Gulbarga           &  5 & 196.35 &  5.23 & Saran              &  4 & 220.00 & 11.74 \\
Guntur             &  6 & 215.19 & 11.27 & Sarguja            &  5 & 205.93 &  7.88 \\
Hailakandi         &  5 & 211.30 & 20.53 & Shamli             &  2 & 205.03 & 19.23 \\
Hamirpur           &  7 & 206.05 & 12.28 & Shimla             &  5 & 201.51 & 17.14 \\
Hyderabad          &  5 & 201.62 & 19.61 & Shimoga            &  4 & 206.49 & 11.11 \\
ImphalWest         &  8 & 217.78 &  3.38 & Sindhudurg         &  3 & 199.16 & 11.30 \\
Jahanabad          &  6 & 217.34 & 10.13 & Sitamarhi          &  5 & 218.35 & 11.14 \\
Jaipur             &  5 & 220.00 & 19.31 & Solapur            &  6 & 218.75 &  9.78 \\
Jaisalmer          &  3 &  47.50 &  0.46 & Sonitpur           &  6 & 231.29 & 17.15 \\
Jalaun             &  6 & 220.00 & 12.19 & SouthGarohills     &  2 & 227.13 & 21.99 \\
Jalpaiguri         &  4 & 217.37 &  8.42 & SouthTripura       &  3 &   8.87 &  0.66 \\
Jamtara            &  6 & 209.42 & 16.17 & Srikakulam         &  3 & 230.30 & 10.88 \\
Jamui              &  5 & 208.42 & 10.30 & Srinagar           &  4 & 110.93 &  0.00 \\
Jashpur            &  7 & 205.37 &  7.72 & SriSatyaSai        &  4 & 206.86 & 25.91 \\
Jhajjar            &  4 & 200.26 & 15.08 & Sukma              &  7 & 135.28 &  8.16 \\
Jhargram           &  3 & 199.75 &  9.80 & Supaul             &  5 & 202.81 &  9.00 \\
JyotibaPhuleNagar  &  5 & 210.93 & 14.10 & TehriGarhwal       &  5 & 207.01 & 14.03 \\
Kabirdham          &  4 & 220.00 &  9.10 & Thiruvananthapuram &  3 &  17.47 &  2.25 \\
Kaimur             &  5 & 174.54 &  4.33 & Umaria             &  4 & 146.89 &  6.23 \\
KamrupMetropolitan &  4 & 218.90 & 21.38 & Unakoti            &  5 & 211.59 & 22.55 \\
Kanyakumari        &  2 & 203.54 &  7.55 & Uttarkashi         &  6 & 207.45 & 10.92 \\
Kapurthala         &  4 & 196.89 & 22.56 & Vaishali           &  6 & 220.00 & 13.44 \\
KarbiAnglong       &  6 &  89.93 &  0.94 & Valsad             &  5 & 200.00 & 21.38 \\
Karimnagar         &  5 & 204.37 &  7.76 & Varanasi           &  5 & 207.66 &  8.82 \\
Kasaragod          &  2 &  58.04 &  7.41 & Vishakapattanam    &  5 & 185.20 & 11.18 \\
Katihar            & 10 & 164.69 &  9.27 & Washim             &  7 &  79.50 &  6.73 \\
Katni              &  6 & 220.00 & 20.14 & Wayanad            &  3 &  22.42 &  2.64 \\
Khordha            &  5 & 215.44 & 23.85 & WestChamparan      &  8 & 203.17 & 18.01 \\
Kishanganj         &  4 & 197.64 &  9.81 & WestGaroHills      &  7 & 200.00 & 21.15 \\
Kohima             & 15 & 220.00 & 15.14 & WestTripura        &  5 & 208.97 & 21.17 \\
Kolkata            &  2 & 220.00 & 11.85 &                    &    &        &       \\
\end{longtable}

\newpage

%==============================================================
% Table 2: Language-wise distribution
%==============================================================

\begin{longtable}{N C D D | N C D D}
\caption{Audio collection and transcription duration per language (hours), aggregated across all districts.}
\label{tab:languages} \\
\toprule
\textbf{Language} & \textbf{Distr.} & \textbf{Audio} & \textbf{Transc.} & \textbf{Language} & \textbf{Distr.} & \textbf{Audio} & \textbf{Transc.} \\
                  &                 & \textbf{(hrs)} & \textbf{(hrs)}   &                  &                 & \textbf{(hrs)} & \textbf{(hrs)}   \\
\midrule
\endfirsthead

\multicolumn{8}{c}{\tablename\ \thetable\ -- \textit{continued from previous page}} \\
\toprule
\textbf{Language} & \textbf{Distr.} & \textbf{Audio} & \textbf{Transc.} & \textbf{Language} & \textbf{Distr.} & \textbf{Audio} & \textbf{Transc.} \\
                  &                 & \textbf{(hrs)} & \textbf{(hrs)}   &                  &                 & \textbf{(hrs)} & \textbf{(hrs)}   \\
\midrule
\endhead

\midrule \multicolumn{8}{r}{\textit{continued on next page}} \\
\endfoot

\midrule
\multicolumn{8}{r}{\textbf{Grand Total --- Audio: 31255.45 hrs \quad Transcription: 2043.31 hrs}} \\
\bottomrule
\endlastfoot

Hindi         & 153 & 14868.49 & 957.36 & Lepcha     & 1 & 7.58 & 0.00 \\
Bengali       &  67 &  2377.73 & 154.51 & Desia      & 1 & 7.38 & 0.00 \\
Telugu        &  27 &  2339.10 & 158.05 & Santali    & 3 & 7.05 & 0.49 \\
Kannada       &  18 &  2245.43 & 151.04 & Bearybashe & 2 & 6.96 & 0.18 \\
Marathi       &  53 &  1124.76 &  69.16 & Khandeshi  & 1 & 5.66 & 0.00 \\
Tamil         &  11 &   850.84 &  24.38 & Nyishi     & 1 & 5.20 & 0.11 \\
Odia          &  11 &   593.36 &  51.77 & Chakhesang & 3 & 4.71 & 0.39 \\
Chakma        &   3 &   485.75 &  50.28 & Ao         & 2 & 4.69 & 0.49 \\
Bhojpuri      &  26 &   482.00 &  25.20 & Rengma     & 1 & 4.36 & 0.51 \\
Garo          &   3 &   471.00 &  47.25 & Sindhi     & 5 & 3.97 & 0.05 \\
Maithili      &  29 &   469.35 &  21.25 & Rongmei    & 3 & 3.94 & 0.29 \\
Nepali        &   6 &   420.19 &  38.17 & Bhili      & 2 & 3.06 & 0.16 \\
Chhattisgarhi &  19 &   408.71 &  19.39 & Bagheli    & 3 & 2.92 & 0.13 \\
Assamese      &   9 &   357.28 &  29.65 & Koya       & 1 & 2.64 & 0.00 \\
Malayalam     &  16 &   352.06 &  41.86 & Sikkimese  & 1 & 2.54 & 0.00 \\
English       &  76 &   328.38 &  23.96 & Jaipuri    & 2 & 2.31 & 0.20 \\
Gujarati      &   8 &   297.11 &  24.01 & Tangkhul   & 1 & 2.15 & 0.30 \\
Nagamese      &   4 &   275.36 &  23.78 & Tagin      & 1 & 1.99 & 0.12 \\
Punjabi       &   9 &   236.70 &  21.72 & Bhatri     & 3 & 1.71 & 0.07 \\
Manipuri      &   3 &   207.17 &   0.45 & Powari     & 1 & 1.50 & 0.13 \\
Mizo          &   1 &   201.50 &  20.75 & Malvi      & 1 & 1.38 & 0.00 \\
Rajasthani    &   7 &   189.56 &  12.08 & Kurmali    & 2 & 1.36 & 0.07 \\
Urdu          &  41 &   173.10 &   2.18 & Dorli      & 1 & 1.32 & 0.11 \\
Marwari       &   7 &   159.46 &   9.04 & Pahadi     & 1 & 1.25 & 0.12 \\
Garhwali      &   2 &   136.80 &   8.07 & Yimchunger & 2 & 1.18 & 0.12 \\
Wancho        &   2 &   121.40 &  11.91 & Shekhawati & 1 & 1.08 & 0.08 \\
Magahi        &  12 &   111.42 &   6.66 & Sangtam    & 2 & 0.97 & 0.12 \\
Angika        &  12 &   101.06 &   3.82 & Bagri      & 2 & 0.76 & 0.06 \\
Karbi         &   1 &    85.69 &   0.63 & Thethi     & 2 & 0.73 & 0.04 \\
Bajjika       &   4 &    84.76 &   4.19 & Lambani    & 2 & 0.66 & 0.02 \\
Konkani       &   2 &    67.89 &   3.99 & Sylheti    & 1 & 0.61 & 0.06 \\
Halbi         &   2 &    54.70 &   2.21 & Liangmai   & 3 & 0.55 & 0.04 \\
Kokborok      &   4 &    43.59 &   4.55 & Wagdi      & 1 & 0.54 & 0.03 \\
Tulu          &   1 &    39.60 &   3.05 & Zeme       & 2 & 0.53 & 0.06 \\
Haryanvi      &   4 &    38.53 &   0.65 & Galo       & 1 & 0.51 & 0.05 \\
Sambalpuri    &   2 &    35.29 &   1.33 & Duruwa     & 1 & 0.50 & 0.00 \\
Kashmiri      &   3 &    33.59 &   0.07 & Thadou     & 1 & 0.45 & 0.06 \\
Khortha       &   5 &    32.38 &   3.75 & Hajong     & 1 & 0.40 & 0.04 \\
Kumaoni       &   2 &    28.06 &   1.70 & Dogri      & 2 & 0.35 & 0.00 \\
Sadri         &   4 &    27.46 &   0.95 & Mewati     & 1 & 0.34 & 0.04 \\
Surjapuri     &   1 &    25.78 &   0.25 & Harauti    & 1 & 0.34 & 0.02 \\
Lotha         &   2 &    22.10 &   0.05 & Mewari     & 1 & 0.30 & 0.04 \\
Khariboli     &  10 &    20.10 &   1.40 & Vaiphei    & 1 & 0.25 & 0.02 \\
Surgujia      &   2 &    16.95 &   0.65 & Rajbanshi  & 1 & 0.24 & 0.02 \\
Nimadi        &   1 &    16.12 &   0.00 & Limbu      & 1 & 0.23 & 0.00 \\
Malvani       &   4 &    15.87 &   0.92 & Phom       & 1 & 0.23 & 0.02 \\
Kurukh        &   3 &    11.73 &   0.39 & Sirmauri   & 1 & 0.18 & 0.00 \\
Bundeli       &   5 &    10.31 &   0.83 & Agariya    & 1 & 0.16 & 0.00 \\
Idu Mishmi    &   1 &     9.95 &   1.08 & Baghati    & 1 & 0.08 & 0.00 \\
Angami        &   2 &     9.77 &   0.17 & Paniya     & 1 & 0.07 & 0.01 \\
Sumi          &   3 &     9.76 &   1.54 & Mara       & 1 & 0.05 & 0.00 \\
Gondi         &   3 &     9.58 &   0.29 & Kuki       & 1 & 0.03 & 0.00 \\
Awadhi        &   9 &     8.85 &   0.27 &            &   &      &      \\
\end{longtable}

\begin{comment}
\section{Technical appendices and supplementary material}
Technical appendices with additional results, figures, graphs, and proofs may be submitted with the paper submission before the full submission deadline (see above). You can upload a ZIP file for videos or code, but do not upload a separate PDF file for the appendix. There is no page limit for the technical appendices. 

Note: Think of the appendix as ``optional reading'' for reviewers. The paper must be able to stand alone without the appendix; for example, adding critical experiments that support the main claims to an appendix is inappropriate. 
\end{comment}
%%%%%%%%%%%%%%%%%%%%%%%%%%%%%%%%%%%%%%%%%%%%%%%%%%%%%%%%%%%%

%\newpage
%\input{checklist.tex}

\end{document}